\documentclass{article}%
\usepackage{amsmath}
\usepackage{amsfonts}
\usepackage{amssymb}
\usepackage{graphicx}%
\newcommand{\scetI}{ SCET$_{\text{I}}$ }
\newcommand{\scetII}{ SCET$_{\text{II}}$ }
\newcommand{\Lamb}{\bar{\Lambda} }

\newcommand{\xiB}{\xi_{\pi}^{B}}
\def\Dirac#1{#1\hskip-6pt/}
\def\slash#1{#1\hskip-4pt/}

\begin{document}

\rightline{RUB-TPII-06/07}

\vspace*{1cm}

\begin{center}

{\Large  Radiative corrections to hard spectator
scattering in $B\rightarrow \pi\pi$  decays }\\[0.5cm]

\vspace*{1cm}

\renewcommand{\thefootnote}{\fnsymbol{footnote}}
 Nikolai~Kivel\footnote{
On leave of absence from St.~Petersburg Nuclear Physics Institute,
188350, Gatchina, Russia
} \\[3mm]
{ \it Institut f\"ur Theoretische Physik II,Ruhr-Universit\"at
Bochum, D-44780 Bochum, Germany
 }
\vspace*{1cm}

\end{center}
\renewcommand{\thefootnote}{\arabic{footnote}}

\begin{abstract}
We present the calculation of the next-to-leading corrections to
the  tree amplitudes which appear in the description of
non-leptonic $B$-decays in the factorization approach. These
corrections, together with radiative corrections to the jet
functions, represent the full next-to-leading contributions to the
dominant hard spectator scattering term generated by operators
$O_{1,2}$ in the decay amplitudes. Using obtained analytical
results we  estimate $B\rightarrow\pi\pi$ branchings fractions in
the physical ( or BBNS) factorization scheme. We have also found
that the imaginary part generated in the hard spectator scattering
term is rather large compared to the imaginary part of the vertex
contribution.
\end{abstract}
\bigskip

\newpage

\section*{Introduction}

The processes of non-leptonic decays of $B~$ mesons are considered
as one of the most interesting topics at present time. They  are
sensitive to the physics of the standard model and provide a nice
possibility to search for new physics effects. The channel of two
$\pi$-mesons is very interesting because phenomenological analysis
of the corresponding branchings and $CP$ asymmetries can be done
to a good accuracy in a model independent way \cite{London}. Last
few years $B\rightarrow\pi\pi$ branching fractions and
$CP$-asymmetries have been measured by BABAR \cite{BABAR} and
BELLE \cite{BELLE} collaborations.

Important progress has also been achieved in the theory.
There was
suggested a new approach which is based on the idea of QCD factorization.
The
factorization allows, in some sense, to constrain the strong
interaction background in a model independent way and therefore
provides a theoretical basis for analysis of $B-$decays
which can be considered as an alternative  to the
traditional phenomenological fits.

The factorization theorem for nonleponic decays has been initially
suggested in \cite{BBNSPRL}. The statement has been proved by
explicit calculations at the leading and next-to-leading orders.
The general proof of the factorization to all orders can be done
using the so-called soft-collinear effective theory
(SCET)\cite{SCET}. The application of SCET technique to the two mesons
decays has been formulated in \cite{SCETB2pi,SCETB2PP}. The
presence of two hard scales $\mu\sim m_{b}$ and
$\mu\sim\sqrt{\Lamb m_{b}}$ leads to two steps matching
QCD$\rightarrow$\scetI$\rightarrow$\scetII
 with corresponding two independent coefficient functions which can be
 calculated  systematically in
perturbative QCD. The non-perturbative dynamics is described
 by the matrix elements of the  light-cone operators constructed from the fields
 of the  \scetII  effective theory.
These unknown functions are universal for all processes and
therefore can be constrained from the global analysis.
Phenomenological applications to the nonleptonic decays
 based on the factorization  have been
considered in several papers.  The QCD factorization approach  (the
so-called BBNS or physical scheme) was used in \cite{BBNS}.
A different  analysis  on basis of SCET  was suggested in
\cite{SCETB2PP,Williamson:2006hb}. Although both approaches are based on the same
theoretical idea they are different in the consideration of some
phenomenological moments, see for instance discussion in
\cite{BBNScom}.

 An important question which appears in application of the
factorization is applicability of perturbation theory at
relatively moderate scale
$\mu\sim\sqrt{\Lamb m_{b}}\sim1-2$ GeV . This situation
arises at the second step of the matching
\scetI$\rightarrow$\scetII. In order to  answer this question, the
next-to-leading calculations of the so called jet coefficient
functions have been done in \cite{BHjet,BY05,Krn}. It was
demonstrated \cite{BY05} that the radiative corrections are large
but, on the other side, do not indicate any problem for the
applicability of the perturbative expansion. But the full
next-to-leading contributions also include  corrections to the hard
coefficient functions which describe matching QCD to \scetI
effective theory. A priory, such corrections could also be
considered as a source of quite large contributions, especially
for the color suppressed amplitudes in the BBNS analysis \cite{BY05}.
Therefore the second tail of the next-to-leading corrections,
corresponding to the matching of QCD to \scetI  at $\mu\sim m_{b}$
also have to be computed. An other important motivation for such
a calculation is the observation that the imaginary part of  hard
spectator amplitude arises only from the
radiative corrections. If it can produce sizable corrections to the
$CP$--asymmetries then such contribution is very important for the
phenomenological analysis.

Recently, such calculations have been carried out and results
are presented in \cite{Beneke05} for
the graphical tree amplitudes and for the penguin amplitudes
\cite{Beneke06}. In this paper we present the calculations of the
radiative corrections to the graphical tree amplitudes. Our
results have been computed using different technical approach and
can be considered as an independent derivation of the corresponding
corrections.

Our paper has the following structure. In Section~I we introduce the
basic
notation and review, for convenience,
 the formulation of the factorization theorem for
  $B\rightarrow\pi\pi$ decays.
In Section II we discuss  the matching from QCD to \scetI.
We define the basic set of \scetI operators and recalculate the
leading order coefficient functions. The calculation of the one loop
diagrams and results for the coefficient functions are given in
Section III. Section IV is devoted to the numerical estimates of
the branching fractions. The discussion of some technical questions
and the analytical results for the individual diagrams are
presented in the Appendix.

\bigskip

\section{QCD factorization for $B\rightarrow\pi\pi$
decays }

For the convenience  we review shortly the basic QCD factorization approach
suggested in \cite{BBNS}.
The amplitudes of two pion  decays are given by matrix elements
\begin{equation}
A_{\pi\pi}=\left\langle \pi(p^{\prime})\pi(p)\left\vert H_{eff}\right\vert
\bar {B}(P)\right\rangle \label{Apipi}%
\end{equation}
with the effective Hamiltonian%
\begin{align}
H_{eff}  &  =\frac{G_{F}}{\sqrt{2}}\lambda_{u}^{(d)}\left(  C_{1}O_{1}%
^{u}+C_{2}O_{2}^{u}+\sum_{i=3}^{10}C_{i}O_{i}+C_{7\gamma}O_{7\gamma}%
+C_{8g}O_{8g}\right)  +h.c.\\
&  +\frac{G_{F}}{\sqrt{2}}\lambda_{c}^{(d)}\left(  C_{1}O_{1}^{c}+C_{2}%
O_{2}^{c}+\sum_{i=3}^{10}C_{i}O_{i}+C_{7\gamma}O_{7\gamma}+C_{8g}%
O_{8g}\right)  +h.c. \label{Heff}%
\end{align}
where $\lambda_{u}^{(d)}=V_{ub}V_{ud}^{\ast},~\lambda_{c}^{(d)}=V_{cb}%
V_{cd}^{\ast}~$, $C_{i}$ and $O_{i}^{p}$ are coefficient functions and
effective four-fermion operators respectively. In particular,
the explicit expressions for
 the current-current operators are
\begin{align}
O_{1}^{u}  &  =(\overline{u}~b)_{V-A}(\overline{d}u)_{V-A},\ O_{2}%
^{u}=(\overline{u}_{\beta}~b_{\alpha})_{V-A}(\overline{d}_{\alpha}u_{\beta
})_{V-A}\label{O12}\\
O_{1}^{c}  &  =(\overline{c}~b)_{V-A}(\overline{d}c)_{V-A},\ O_{2}%
^{c}=(\overline{c}_{\beta}~b_{\alpha})_{V-A}(\overline{d}_{\alpha}c_{\beta
})_{V-A}%
\end{align}
where as usual $V-A=\gamma^{\mu}(1-\gamma_{5})$ and indices $\alpha,~\beta$
stand for the color. The definitions of the remaining terms are standard and can be
found, for instance, in \cite{Buras}. Taking into account the structure of
the effective Hamiltonian (\ref{Heff}) the decay amplitudes $A_{\pi\pi}$ can be
conveniently rewritten through the effective amplitudes $\alpha_{i}$ in the
following way \cite{BBNS}:\footnote{We have slightly changed the original notation
removing $f_0$ from the normalization factor}%
\begin{equation}
A_{\pi^{+}\pi^{-}}=-\lambda_{u}^{(d)}\frac{iG_{F}}{\sqrt{2}}M_{B}^{2}f_{\pi
}\left[  \alpha_{1}+\alpha_{4}^{u}+\alpha_{4,EW}^{u}\right]  -\lambda
_{c}^{(d)}\frac{iG_{F}}{\sqrt{2}}M_{B}^{2}f_{\pi}\left[  \alpha_{4}^{c}%
+\alpha_{4,EW}^{c}\right],
\end{equation}%
\begin{equation}
A_{\pi^{0}\pi^{0}}=\lambda_{u}^{(d)}\frac{iG_{F}}{\sqrt{2}}M_{B}^{2}f_{\pi
}\left[  -\alpha_{2}+\alpha_{4}^{u}-\frac{3}{2}\alpha_{3,EW}^{u}-\frac{1}%
{2}\alpha_{4,EW}^{u}\right]  +\lambda_{c}^{(d)}\frac{iG_{F}}{\sqrt{2}}%
M_{B}^{2}f_{\pi}\left[  \alpha_{4}^{c}-\frac{3}{2}\alpha_{3,EW}^{u}-\frac
{1}{2}\alpha_{4,EW}^{c}\right],
\end{equation}
where we have neglected the annihilation contributions.
The amplitudes $\alpha_{i}$ describe the matrix elements of the different
operators in (\ref{Heff}). Namely, $\ \alpha_{1,2}$ \ gives the matrix elements
 of the
current-current operators $O_{1,2}$, \ $\alpha_{4}^{u,c}$ and $\alpha
_{3,4,EW}^{u,c}~\ \ $ denote the QCD and Electro-Weak penguin contributions
respectively. From the isospin symmetry one has \
\begin{equation}
\sqrt{2}A_{\pi^{0}\pi^{-}}=A_{\pi^{0}\pi^{0}}+A_{\pi^{+}\pi^{-}}%
\end{equation}

 We used notation $M_{B}$
for $B-$meson mass, $f_{\pi}$ is pion decay constant and~below $f_{0}\equiv
f_{0}(0)=f_{+}(0)$ denotes $B\rightarrow\pi$ transition
form factors at $q^2=0$:
\begin{equation}
\left\langle \pi(p)\left\vert \overline{q}\gamma^{\mu}b\right\vert
\bar {B}(P)\right\rangle =f_{+}(q^{2})\left[  P^{\mu}+p^{\mu}-\frac
{M_{B}^{2}-m_{\pi}^{2}}{q^{2}}q^{\mu}\right]  +f_{0}(q^{2})\frac{M_{B}%
^{2}-m_{\pi}^{2}}{q^{2}}q^{\mu}\, .%
\end{equation}
 The amplitudes $\alpha_{i}^{p}$ include all dynamical
information about the
decays. In the limit of large $b-$quark mass \ $m_{b}\rightarrow\infty$ the QCD
factorization approach makes it possible to calculate amplitudes $\alpha_{i}^{p}$
to the leading power accuracy. Let us consider the matrix elements
$\alpha_{1,2}$ which provide dominant contribution to the branching fractions.
Their expressions are given by
\begin{align}
\alpha_{i}  &  =f_{0}\int_{0}^{1}du~V_{i}(u)\varphi_{\pi}(u)+\int_{0}%
^{1}du~\varphi_{\pi}(u)\int_{0}^{1}dz~T_{i}(u,z)~\xi_{\pi}^{B1}%
(z),\label{alphai}\\
\xi_{\pi}^{B1}(z)  &  =f_{B}f_{\pi}\int_{0}^{\infty}d\omega\int_{0}%
^{1}dx~~\phi_{B}(\omega)J(z,x,\omega)~\varphi_{\pi}(x) \label{xiB}%
\end{align}
where functions $V_{i}$ and $T_{i}$ are the hard coefficient functions which
can be computed in the perturbation theory order by order in QCD
coupling~$\alpha_{S}$:\footnote{In this paper we always assume that
perturbative
expansion of any quantity $R$ is defined as $R=R^{LO}+\frac{\alpha_{S}}{2\pi
}R^{NLO}+...$}
\begin{align}
V_{i}  &  =V_{i}^{LO}(u)+\frac{\alpha_{S}}{2\pi}V_{i}^{NLO}%
(u,z,m_{b}/\mu_F)+...\label{Vi:PTexp}\\
T_{i}  &  =T_{i}^{LO}(u)+\frac{\alpha_{S}}{2\pi}T_{i}^{NLO}%
(u,z,m_{b}/\mu_F)+... \label{Ti:PTexp}%
\end{align}
 The hard coefficient functions
describe the hard subprocess in which quarks and gluons are highly virtual, with
typical hard momenta  $p_{h}^{2}\sim m_{b}^{2}$ . Performing integration
over such fluctuations we reduce QCD to the effective theory SCET$_{\text{I}}$
which however still contains large  \ hard-collinear fluctuations
of order $p_{hc}^{2}\sim m_{b}\Lamb .$ Integrating over these degrees of freedom
we reduce SCET$_{\text{I}}~$\ to the low energy effective theory
SCET$_{\text{II}}$ which contains only collinear and soft particles with small
 $p_{c}^{2}\sim p_{s}^{2}\sim\Lamb^{2}$ off-shell momenta. The coefficient
function which appears at this step is the so-called jet-function
$J(z,x,\omega)$ :%
\begin{equation}
J(z,x,\omega)=\alpha_{S}(\mu_{hc})~J^{LO}(z,x,\omega)+...
\end{equation}
where the hard-collinear scale $\mu_{hc}\sim\sqrt{m_{b}\Lamb}.$ The soft physics
is encoded by the matrix elements of SCET$_{\text{II}}$ operators
constructed from the soft and collinear fields. These matrix elements are
parametrised by non-perturbative light-cone distribution amplitudes (LCDA)
$\varphi_{\pi},\phi_{B}$ \ and decay constants $f_{\pi},~f_{B}$. Their
explicit definitions are given by
\begin{equation}
f_{\pi}\varphi_{\pi}(x)=i\int\frac{d\lambda}{\pi}e^{-i(2x-1)(p.n)\lambda
}\left\langle \pi^{-}(p)\left\vert \bar{d}(\lambda n)~\Dirac {n}~\gamma
_{5}~u(-\lambda n)\right\vert 0\right\rangle , \label{pionDA}%
\end{equation}
where $\bar n$ and $n$ are the light cone vectors: $n^{2}=\bar n^{2}=0,(n\cdot \bar n)=2~$
and pion decay constant defined as%
\begin{equation}
\left\langle \pi^{-}(p)\left\vert \overline{d}(0)~\Dirac{n}\gamma_{5}%
~u(0)\right\vert 0\right\rangle =-if_{\pi}(p.n)\, ,
\end{equation}
that implies $\int dx~\varphi_{\pi}(x)=1$.
$B-$meson LCDA is  given by
\begin{equation}
F_{stat}(\mu)\sqrt{M_{B}}\phi_{B}(\omega)=-i\int\frac{d\lambda}{2\pi
}e^{+i\omega\lambda}\left\langle 0\left\vert \bar{q}(\lambda n)\Dirac
{n}\gamma_{5}h_{v}(0)\right\vert \bar {B}(P)\right\rangle ,
\label{phiB}%
\end{equation}
where $v=\frac{1}{2}(\bar n+n)$ is the velocity of $B-$meson\ at the rest frame.
The $M_{B}$-independent decay constant $F_{stat}(\mu)$ \cite{Ji:1991pr}
is defined as%
\begin{equation}
F_{stat}(\mu)=\sqrt{M_{B}}~f_{B}/K_F(\mu),~K_F(\mu)=1+\frac{\alpha_{S}C_{F}%
}{4\pi}\left(  3\ln\frac{m_{b}}{\mu}-2\right)
\label{Fstat}
\end{equation}
where the physical decay constants $f_{B}$ is given by%
\begin{equation}
\left\langle 0\left\vert \overline{q}\gamma^{\mu}\gamma_{5}b\right\vert
\bar {B}(P)\right\rangle =if_{B}M_{B}v^{\mu}.
\end{equation}
As it was shown in \cite{Grozin96,Braun03}, the normalization integral
for $\phi_{B}(\omega)$ and higher moments
are not defined   and  therefore the non-local matrix element
(\ref{phiB})
can not be reduced to the local
matrix element (\ref{Fstat}). This feature makes this function quite different
from the standard LCDA 's of light mesons.

As we can see from  equation (\ref{alphai}) the jet function appears in the
second term only. This term describes the hard spectator interaction.
 For the fist term in
(\ref{alphai}) the matching to SCET$_{\text{II}}$ \ is not
possible due to the overlap of the soft and collinear regions, see
discussions \cite{Lange03,Beneke03}.
 Such contribution is known as soft-overlap
form factor. In the BBNS prescription this form factor is excluded using
 the so-called "physical scheme". In this
approach the soft-overlap form factor is rewritten as a sum of  physical form
factor $f_{0}$ and of hard spectator scattering contribution
(the details are given  below in the text).

Explicit expressions for the hard coefficient functions read
\cite{BBNS}
$(i\pm1\equiv i+(-1)^{i+1})$%
\begin{align}
V_{i}(u)  &  =\left(  C_{i}+\frac{C_{i\pm1}}{N_{c}}\right)  +\frac{\alpha_{S}%
}{2\pi}~V_{i}^{NLO}(u)+O(\alpha_{S}^{2}),\label{Vi}\\
V_{i}^{NLO}(u)  &  =\frac{C_{_{i\pm1}}}{N_{c}}V(u),~~\label{ViNLO}\\
V(u)  &  =\frac{1}{2}C_{F}\left(  12\ln\frac{m_{b}}{\mu_{h}}-18+3\left(
\frac{1-2u}{1-u}\ln u-i\pi\right)  +\right. \label{V(u)}\\
&  \left.  \left[  2\text{Li}_{2}[u]-\ln^{2}u+\frac{2\ln u}{1-u}-(3+2i\pi)\ln
u-(u\leftrightarrow\overline{u})\right]  \right)  .
\end{align}%
\begin{equation}
T_{i}(u,z)=-\frac{C_{i\pm1}}{N_{c}}\frac{1}{1-u}+\frac{\alpha_{S}}{2\pi}%
T_{i}^{NLO}(u,z)+O(\alpha_{S}^{2}), \label{Ti}%
\end{equation}
For the jet function in our notation we have%
\begin{equation}
J(z,x,\omega)=\alpha_{S}\left\{
\left[-\pi\frac{C_{F}}{N_{c}}\frac{\delta
(x-z)}{\overline{x}~\omega m_{b}}\right] +
\frac{\alpha_{S}}{2\pi} J^{NLO}(z,x,\omega)+ ...
\right\},
\label{Jet}%
\end{equation}
The next-to-leading order expression for $J^{NLO}(z,x,\omega)$ has been recently
obtained in several papers  \cite{BHjet,BY05,Krn} and we shall not present \
it here.

\section{ Calculation of the hard coefficient functions}

The aim of this section is to discuss some details
relevant for our calculation. We shall reproduce the leading order
results for  hard coefficient functions $T_{i}$
quoted in eq.(\ref{Ti}).\ As it
was discussed in the previous section, $T_i$ is associated  with \ matching QCD
to the effective theory \scetI. Technical details of such
calculations have already been discussed in \cite{SCET1,NBH,BKY04} for the case of
heavy-to-light currents.

First, let us fix the basis of  \scetI operators relevant for
our case. We introduce two  operators with approprite flavor $q$
and chiral structure:
\begin{align}
J^{(A0)}(s)  &  =(\bar{q}_n W_{c})_{s}\left(  1-\frac{i\overleftarrow
{\Dirac \partial}}{i~\bar n\overleftarrow{\partial}}\right)  h_{v},\label{JA0}\\
J^{(B1)}(s)  &  =\frac{1}{m_{b}}(\bar{q}_n W_{c})_{s}(W_{c}^{\dagger}%
i\Dirac{D}_{\perp c}W_{c})_{(-s)}(1-\gamma_{5})h_{v}.~ \label{JB1}%
\end{align}
where we have accepted the notation introduced in  \cite{BCDF}.
The light quarks are
supposed to be collinear fields in SCET approach $ \Dirac n q_n(x)=0$,
 $h_{v}$ is HQET field.
 Notation $W_{c}$ is used for
the hard-collinear Wilson line involving only $\bar n \cdot A_{c}$ component of the
collinear gluon field:%
\begin{equation}
\left( \bar{q}_n W_{c}\right)  _{s}=
\bar {q}_n(s~\bar n)P\exp\left\{  ig\int_{-\infty}^{0}du~\bar n%
A_{c}[(s+u)~\bar n]\right\}  .
\end{equation}
Matrix elements of \ these operators between physical particles define two
\scetI form factors:%
\begin{equation}
\left\langle ~\pi^{+}(p)\left\vert (\bar{d}_n W_{c})~h_{v}\right\vert
\bar {B}(P)\right\rangle =~m_{b}~\xi_{\pi}, \label{xi:pi}%
\end{equation}%
\begin{equation}
\left\langle ~\pi^{+}(p)\left\vert J^{(B1)}(s)\right\vert \bar {B}(P)%
\right\rangle =m_{b}\int_{0}^{1}dz~e^{is(2z-1)m_{b}}~\xi_{\pi}^{B1}(z),
\label{xi:B1}%
\end{equation}
where  dependence of the form factors on mass $m_{b}$ is
implied.
In order to obtain the factorization
formula (\ref{alphai}) one has to perform matching of the effective
Hamiltonian (\ref{Heff}) to the operators in SCET$_{\text{I}}$ (\ref{JA0}) and
(\ref{JB1}) . We shall focus our attention on the contributions of the
current-current operators $O_{1,2}$ because they provide \ dominant part of
the two body decay amplitude.
Then for the \ matrix element $A_{\pi\pi}$ (\ref{Apipi}) we
obtain
\begin{align}
A_{\pi\pi}^{i}  &  =\frac{G_{F}}{\sqrt{2}}\lambda_{u}^{(d)}\left\langle
\left(  \pi\pi\right)  _{i}\left\vert C_{1}O_{1}^{u}+C_{2}O_{2}^{u}\right\vert
\bar {B}\right\rangle =-\frac{iG_{F}}{\sqrt{2}}M_{B}^{2}f_{\pi}%
(~\alpha_{i}~),\label{Apipi1}\\
\alpha_{i}  &  =\left(  ~\xi_{\pi}~v_{i}\ast\varphi_{\pi}+\varphi_{\pi}\ast
t_{i}\ast\xi_{\pi}^{B1}\right)  ,
\end{align}
where $v_{i}~$\ and $t_{i}$ denote hard coefficient functions and by asterisk
$\ast$ we denote, for simplicity, convolution integrals. The index $i$
is introduced to distinguish
two possible final states $\left(  \pi\pi\right)  _{i=1}=\pi^{+}\pi^{-}$,
$\left(  \pi\pi\right)  _{i=2}=\pi^{0}\pi^{0}$. Corresponding matrix elements
define
amplitudes $\alpha_{1}$ and $\alpha_{2}$ respectively.

In the physical scheme one has to express \scetI form factor $~\xi_{\pi}$ through the physical
form factor $f_{+}(0)=f_{0}$ \cite{BY05}:%
\begin{equation}
\xi_{\pi}=\frac{1}{C^{A0}}f_{0}-\frac{1}{C^{A0}}C^{B1}\ast\xi_{\pi}^{B1},
\label{xi:to:f0}%
\end{equation}
where $C^{A0}$ and \ $C_{+}^{B1}$ are the hard coefficient functions which
appear in matching of \ the scalar heavy-light QCD current to the
 operators (\ref{JA0}, \ref{JB1}):%
\begin{equation}
\overline{q}~b=C^{A0}\ast J^{A0}+C^{B1}\ast J^{B1}+O(1/m_{b})
\end{equation}

Inserting equation (\ref{xi:to:f0}) into  (\ref{Apipi1}) we obtain%
\begin{equation}
\alpha_{i}=f_{0}~v_{i}/C^{A0}\ast\varphi_{\pi}+\varphi_{\pi}\ast\left(
t_{i}-v_{i}~C^{B1}/C^{A0}\right)  \ast\xi_{\pi}^{B1}, \label{Apipi:2}%
\end{equation}
Comparing this expression with equation (\ref{alphai}) we find
\begin{align}
V_{i}(u)  &  =v_{i}(u)/C^{A0},\label{Vi:vi}\\
T_{i}(u,z)  &  =t_{i}(u,z)-v_{i}~(u)~C^{B1}(z)/C^{A0}~. \label{Ti:ti}%
\end{align}
These expressions define precisely the coefficient functions $V_{i}$ and $T_{i}$
in the physical scheme through the matching coefficient $v_{i}$ and $t_{i}$ of
the effective operators $O_{1,2}$. Introducing perturbative expansions for
the coefficient functions:%
\begin{align}
C^{A0}  &  =1+\frac{\alpha_{S~}}{2\pi}C_{NLO}^{A0}+...,\label{CA0}\\
C_{+}^{B1}  &  =C_{LO}^{B1}+\frac{\alpha_{S~}}{2\pi}C_{NLO}^{B1}+...,
\label{CB1}%
\\
v_{i}(u)  &  =v_{i}^{LO}(u)+\frac{\alpha_{S}}{2\pi}v_{i}^{NLO}(u)+...~,\\
t_{i}(u,z)  &  =t_{i}^{LO}(u,z)+\frac{\alpha_{S}}{2\pi}t_{i}^{NLO}(u,z)+...~.
\end{align}
we obtained for the functions $V_{i}$ and $T_{i}$ in (\ref{Vi:PTexp}) and
(\ref{Ti:PTexp})
\begin{align}
V_{i}^{LO}  &  =v_{i}^{LO},\label{VLOi:1}\\
V_{i}^{NLO}  &  =v_{i}^{NLO}-v_{i}^{LO}C_{NLO}^{A0}~, \label{VNLOi:2}%
\\
~T_{i}^{LO}  &  =\left(  t_{i}^{LO}-v_{i}^{LO}C_{LO}^{B1}\right)
,\label{TLO:1}\\
T_{i}^{NLO}  &  =(t_{i}^{NLO}-v_{i}^{LO}~C_{NLO}^{B1})-C_{LO}^{B1}~V_{i}%
^{NLO}. \label{TNLO:2}%
\end{align}
One can  observe that \ subtraction terms
$-v_{i}^{LO}C_{NLO}^{A0}$ in equation (\ref{VNLOi:2}),
$-v_{i}^{LO}C_{LO}^{B1}$ and $-v_{i}^{LO}~C_{NLO}^{B1}$ in
(\ref{TLO:1}) and (\ref{TNLO:2}) can cancel the contributions \ of
the "factorizable" diagrams which can be considered as appropriate
product of the two matrix elements:%
\begin{equation}
\left\langle \left(  \pi\pi\right)  _{i}\left\vert O_{1,2}^{u}%
\right\vert \bar {B}\right\rangle _{\text{fact}}\sim\left\langle
\pi\left\vert \overline{q}\Gamma q\right\vert 0\right\rangle ~\left\langle
\pi\left\vert \overline{q}\Gamma b\right\vert \bar {B}\right\rangle
\label{fact}%
\end{equation}
If it is fulfilled, then such diagrams can be ignored in the
calculations of the coefficient functions $T_{i}^{LO}$ and
$T_{i}^{NLO}$.

In our paper we shall obtain the coefficient functions $T_{i}$
computing  matrix elements with quarks and gluons. For that
purpose we define the  perturbative analogs of the discussed form factor
$\xi_{\pi}^{B1}$ and LCDA $\varphi_{\pi}$. To define \scetI form factor
let us consider as external state a hard-collinear quark and a gluon \ with momenta
$p_{1}$ and $p_{2}$ respectively:
\begin{align}
p_{1}  &  =z~p+(p_{1}n)\frac{\bar n}{2}+p_{1\perp},\\
p_{2}  &  =\bar{z}~p+(p_{2}n)\frac{\bar n}{2}+p_{2\perp},
\end{align}
where $p=m_{b}\frac{n}{2},~\bar{z}=1-z$ and the other components are
$(n\cdot p_{1,2})\sim\Lamb,~~\left\vert p_{1,2\perp}\right\vert \sim\sqrt{\Lamb
m_{b}}$ as it necessary for the hard-collinear momenta. Then we define%
\begin{equation}
\left\langle ~q(p_{1})g(p_{2})\left\vert J^{(B1)}(s)\right\vert h_{v}%
\right\rangle =m_{b}\int_{0}^{1}dz~e^{is(2z-1)(p.\bar
n)}\xi_{B}~\theta(z).~
\label{thetaPT1}%
\end{equation}
 The factor $\xi_{B}$ denotes the relevant combination of the quark spinors and
 gluon polarization vector:
 $~\xi_{B}=\frac{g}{m_{b}^{2}}\bar{q}\slash{e}%
_{g}(1-\gamma_{5})h_{v}$\footnote{Symbols $\bar{q}$ and $h_{v}$ denote
the quark spinors in this formula.}.
We assume that the final gluon is
transversely polarized with the polarization vector $e_{g}$, the color
indices are not shown for simplicity. Performing simple
tree level calculation one finds%
\begin{equation}
~~\theta^{LO}(\tau)=\delta(z-\tau).
\end{equation}
In order to define the \ perturbative analog of the pion LCDA we consider
 quark-antiquark
state with collinear momenta%
\begin{equation}
p_{1}^{\prime}=up^{\prime}+p_{1\bot}^{\prime},~~p_{2}^{\prime}=\bar
{u}p^{\prime}+p_{2\bot}^{\prime},~~p^{\prime}=m_{b}\frac{\bar n}{2}.
\end{equation}
Then\footnote{For simplicity, we do not introduce here the collinear SCET fields
following standard QCD notation.}
\begin{equation}
\left\langle p_{1}^{\prime}p_{2}^{\prime}\left\vert \bar{u}(\lambda
n)~\slash{n}~\gamma_{5}~d(-\lambda n)\right\vert 0\right\rangle =-i~f_{P}%
(p^{\prime}n)\int_{0}^{1}dxe^{i\lambda(2x-1)(p^{\prime}n)}\varphi_{P}(x),
\label{phiP1}%
\end{equation}
where we introduced notation $f_{P}=i~\overline{u}\slash n\gamma
_{5}u/m_{b}$. ~Again, from the leading order calculation one obtains%
\begin{equation}
\varphi_{P}^{LO}(u)=\delta(u-x).
\end{equation}
%

In order to calculate coefficient functions $T_{i}$ \ we introduce
the matrix elements describing the decay of the $b$--quark into  three
quarks and gluon
\begin{align}
\left\langle p_{1}^{\prime}~p_{2}^{\prime},~p_{1}~p_{2}\left\vert C_{1}%
O_{1}+C_{2}O_{2}\right\vert b_{v}\right\rangle _{\text{nf}}-
  \left\langle p_{1}^{\prime},p_{2}^{\prime}~p_{1}p_{2}\left\vert
T\left\{  t_{1}\ast J^{(A0)},\mathcal{L}_{hc}^{int}\right\}  \right\vert
b_{v}\right\rangle_{\text{nf}}
  &  =im_{b}%
^{2}~f_{P}\xi_{B}\ ~\varkappa_{i}^{T}
 \label{me:kappaT}%
\end{align}
and parameterized by form factors $\varkappa_{i}^{T}$ respectively.
\begin{figure}[ptb]
\begin{center}
{
\includegraphics[ height=2.5cm, width=2.7cm
]%
{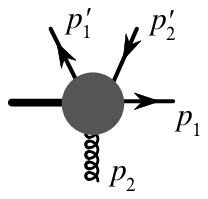}%
}%
\space{\phantom{1cm}}
{
\includegraphics[ height=2.1cm, width=3cm
]%
{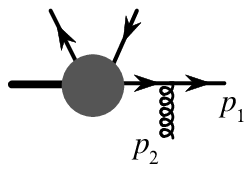}%
}%
\caption{
 Graphical representation of the QCD matrix elements
 which we use to compute $T_i$.
The right graph corresponds to the QCD diagrams with emission
of collinear gluon by  collinear outgoing quark. Such graphs include
contribution from operator $J^{(A0)}$  with quark-gluon vertex
from the leading order SCET lagrangian.
}
\label{LO}%
\end{center}
\end{figure}
By the subscript "nf" we indicate that we exclude the factorizable
diagrams \ (\ref{fact}) which, as expected, cancel in the
transition to the physical scheme. The subtraction term in the left
side of eq.(\ref{me:kappaT}) represents the admixture
of the operator $J^{(A0)}$ with one
 insertion of the interaction vertex
$\mathcal{L}_{hc}^{int}$ from the LO hard-collinear SCET
lagrangian \cite{NBH,BKY04}. Such
contribution describes the emission of collinear gluon from
collinear quark and are present only in the diagrams with topology,
given in Fig.\ref{LO}. Practically this subtraction
can be easily done  by the substitution%
\begin{equation}
\bar{u}(p_{1})\slash e_{g}\frac{i\slash p}{p^{2}}\longrightarrow
\bar{u}_{n}\slash e_{g}\frac{\slash {\bar n} }{2}\frac{i}{(p\bar n)}%
\end{equation}
where $\bar{u}(p_{1})$  is the wave function of
 collinear quark  in full QCD and  $\bar{u}_{n}$ denotes the
hard-collinear spinor in SCET.\

From  the factorization we expect that
\begin{align}
\varkappa_{i}^{T}  &  =\int_{0}^{1}du~\varphi_{P}(u)\int_{0}^{1}%
dz~\widetilde{t}_{i}(u,z)~\theta(z),
\label{kappaT}%
\end{align}
where by tilde we denote coefficient functions of  nonfactorizable diagrams.

\begin{figure}
[ptb]
\begin{center}
\includegraphics[
height=2.5cm, width=5cm
]%
{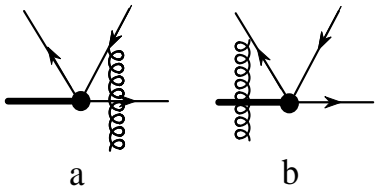}%
\caption{Leading order diagrams for the coefficient functions $T_{i}$ .}%
\label{Fig:LOgraphs}%
\end{center}
\end{figure}

Consider, as example, calculation of $~T_{i}^{LO}.$ Relevant
tree diagrams at the leading order are shown in
Fig.\ref{Fig:LOgraphs}. Two diagrams with emission of a gluon
from the bottom lines represent the factorisable contributions (\ref{fact}) which
cancel against $-v_i^{LO}C_{LO}^{B1}$ in (\ref{TLO:1}).

Straightforward calculation gives (after Fierz transformation)%
\begin{align}
D_{a}  &  =0,~~~\\
D_{b}  &  =i~m_{b~}^{2}f_{P}~\xi_{B}\left(  -\frac{C_{i\pm 1}}{N_{c}}\frac
{1}{\overline{u}}\right)  .
\end{align}
Comparing with eq.(\ref{kappaT}) we obtain:%
\begin{equation}
\int_{0}^{1}du~\varphi_{P}^{LO}(u)\int_{0}^{1}dz~T_{2}^{LO}(u,z)~\theta
^{LO}(z)=\left(  i~m_{b}^{2}~f_{P}\xi_{B}\right)  ^{-1}D_{b}%
\end{equation}
Inserting the leading order expressions for the form factor $\ \theta^{LO}$
and LCDA $\varphi_{P}^{LO}$ we find the leading order hard coefficient
functions:%
\begin{equation}
T_{i}^{LO}(u,z)=-\frac{C_{i\pm1}}{N_{c}}\frac{1}{1-u} \label{TLOi}%
\end{equation}
As one can observe, LO results have no $z-$ dependence.
In the next section we use the same technique to compute the
next-to-leading order corrections.

\section{ Calculation of the coefficient functions $T_{i}$ in the
next-to-leading order}

\begin{figure}
[ptb]
\begin{center}
\includegraphics[
height=2.3263in,
width=5.5979in
]%
{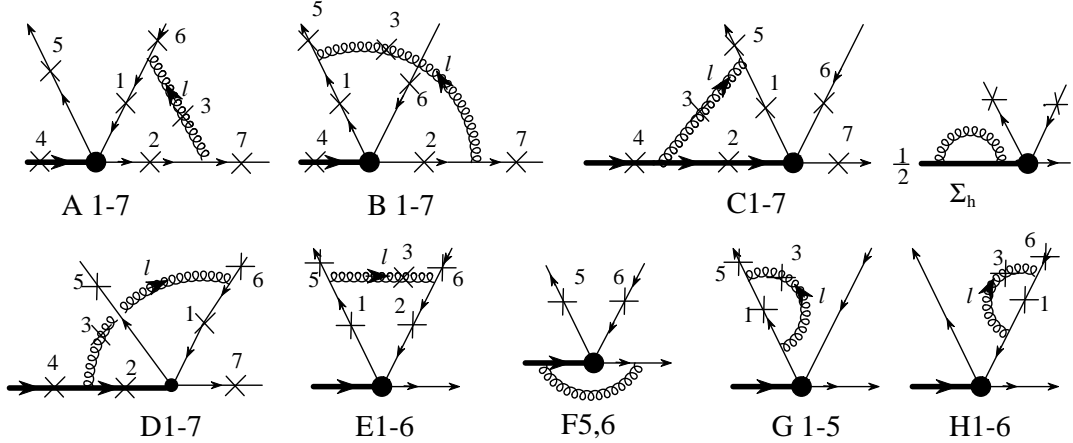}%
\caption{One loop diagrams which have to be computed in QCD. The crossed line
 denotes the emission of the outgoing gluon and the number  gives the index
of the corresponding diagram. The momentum flow is shown only for one line with
the loop momentum $l$ . We skip for simplicity the diagrams with the
light quark wave
function renormalization.}%
\label{Fig:NLOgraphs}%
\end{center}
\end{figure}
Corresponding one-loop diagrams are shown in Fig.\ref{Fig:NLOgraphs}.
Factorisable diagrams, in the sense of (\ref{fact}),
 are not shown for simplicity.
These are the diagrams where the external gluon is emitted from one of
the bottom quark lines and  the
virtual gluon connects only the bottom (upper) quark lines but not upper and bottom.
For the case of form factor $\alpha_1$, these diagrams
 naturally reproduce
subtraction term $v_1^{LO}C_{NLO}^{B1}$ in (\ref{TNLO:1})
and therefore cancel. But analogous situation
 for the $\alpha_2$ is more involved because of the
 different Dirac structure of the operator
vertex. The problem is that corresponding $UV-$ divergent diagrams in
dimensional regularization can not be represented exactly in the factorised
form (\ref{fact}) because Fierz identities can
not be used to regularized diagrams.
Therefore one has to check the exact cancellation against
$v_2^{LO}C_{NLO}^{B1} $ after $UV-$renormalization.
We have fond that in accordance with subtraction scheme, described
below in the text, such cancellation is exact.
Therefore we shall not discuss these
diagrams further.

From the factorization we expect that form factors
$ \varkappa_{i}^{T}$ describing the
matrix elements (\ref{me:kappaT}) of the
renormalized  QCD operators can be represented as a sum of three
contributions:%
\begin{align}
\left(  \varkappa_{i}^{T}~\right)  _{NLO}  &  =\varphi_{P}^{LO}(x^{\prime
})\ast\widetilde{t}_{i}^{NLO}(x^{\prime},z^{\prime})\ast\theta^{LO}(z^{\prime
})
\nonumber\\&
+\varphi_{P}^{NLO}(x^{\prime})\ast\widetilde{t}_{i}^{LO}(x^{\prime
},z^{\prime})\ast\theta^{LO}(z^{\prime})
+\varphi_{P}^{LO}(x^{\prime})\ast\widetilde{t}_{i}^{LO}(x^{\prime
},z^{\prime})\ast\theta^{NLO}(z^{\prime})~ \label{regions}%
\end{align}
where \ $\varphi_{P}^{NLO}$ and $\theta^{NLO}$ denote the contributions of the
  renormalized  matrix elements (\ref{thetaPT1}) and  (\ref{phiP1}) in
the next-to-leading order. The three contributions in (\ref{regions}) can be
associated with four integration regions in the loop integrals. The
\textit{hard }region $k_{i}\sim m_{b},~k^{2}\sim m_{b}^{2}$ provides
contributions to the $\widetilde{t}_{i}^{NLO}(x^{\prime},z^{\prime})$, the
\textit{collinear to} $p^{\prime}$ must be associated with contributions to
$\varphi_{P}^{NLO}(x^{\prime})$, \ the \textit{soft} $\ k_{i}\sim
\Lamb,~k^{2}\sim\Lamb^{2}$ and \textit{collinear to }$p$ regions can be
associated with the $\theta^{NLO}(z^{\prime}).$ \
Substituting in (\ref{regions}) the
explicit expressions for the $\varphi_{P}^{LO},~\theta^{LO}(z^{\prime})$ and
$\widetilde{t}_{i}^{LO}(x^{\prime},z^{\prime})$  we obtain%
\begin{equation}
\widetilde{t}_{i}^{NLO}(u,z)=\left(  \varkappa_{i}^{T}~\right)  _{NLO}%
+\frac{C_{i\pm1}}{N_{c}}
\int_{0}^{1}dx^{\prime}\frac{\varphi_{P}%
^{NLO}(x^{\prime})}{1-x^{\prime}}+
\frac{C_{i\pm1}}{N_{c}}\frac{1}{\bar u}\int_{0}%
^{1}dz^{\prime}\theta^{NLO}(z^{\prime})~.
\end{equation}
Inserting this expression into eq.(\ref{TNLO:2}) and substituting
$C_{LO}^{B1}=-1$ \cite{Chay02,BCDF}\ we find for the NLO coefficient functions in physical scheme:
\begin{equation}
T_{i}^{NLO}(u,z)=\left(  \varkappa_{i}^{T}~\right)  _{NLO}
+\frac{C_{i\pm1}}{N_{c}}
\int_{0}^{1}dx^{\prime}\frac{\varphi_{P}^{NLO}(x^{\prime}%
)}{1-x^{\prime}}+ \frac{C_{i\pm1}}{N_{c}}\frac{1}{\bar u}
\int_{0}^{1}dz^{\prime}\theta
^{NLO}(z^{\prime})~+V_{i}^{NLO}. \label{TNLO}%
\end{equation}

Let us now discuss the calculation of different terms appearing in
(\ref{TNLO}). To perform the calculations of the  diagrams in
 Fig.\ref{Fig:NLOgraphs} one has to
introduce regularization for the ultraviolet ($UV$) and infrared
($IR$) divergencies. We shall use dimensional regularization with
$D=4-2\varepsilon$ to subtract $UV-$divergencies. To compute the
$UV-$divergent subdiagrams of the four-fermion operators we use
$NDR-$scheme with the anticommuting $\gamma_{5}$ matrix. Note that Fierz
identities then can be used only for the renormalized matrix
elements in four dimensions. For the $IR -$divergencies we  use
regularization by off-shell external momenta. Such regularization
makes possible to perform all manipulation with Dirac algebra for
the $UV-$finite integrals in four dimensions.

As an illustration, let us  \ consider a contribution of some
diagram $D_{X}$ which can be
represented in the following way%
\begin{equation}
D_{X}=\int d^{D}l~~\bar{u}\Gamma_{1}b_{v}~\bar{d}\Gamma_{2}u,
\label{DX0}
\end{equation}
where $u,\bar{u},b_{v},\bar{d}$ are quark spinors of given flavor
and matrices $\Gamma_{1}\otimes\Gamma_{2}$ denote some momentum
dependent expressions. Contraction of  spinor indices can be
organized in two different ways which correspond   to the
amplitudes $\alpha_{1}$ or $\alpha_{2}$.
As an example below we consider the  calculation of $\alpha_{2}$.
The same technique also was used  for $\alpha_{1}.$

All graphs can be divided into two groups:  $UV-$divergent and
$UV-$finite. $UV-$divergent subgraphs, which appear in graphs
$G\{1,3,5\}$,$H\{1,3,6\}$ represent usual divergencies of the QCD
Green functions. They are removed by QCD Lagrangian counterterms.
 $UV-$subgraphs in diagrams $A\{4,5,6\}$, $B\{4,5,6\}$,
$C\{4,5,6\}$, $D\{4,5,6\}$ and $F\{5,6\}$ describe renormalization
of the four-fermion operators $O_{1,2}$. Calculation of the
corresponding $UV-$divergent integrals must to be performed in
$D=4-2\varepsilon$. A typical expression for the integrand of
$UV-$divergent graph can be written as
\begin{equation}
\bar{u}\Gamma_{1}b_{v}~\bar{d}\Gamma_{2}u~= N_2^{\mu\nu}
\frac{l_\mu l_\nu}{D[l]}
+N_1^{\mu}\frac{l_\mu}{D[l]}+N_0\frac{1}{D[l]} \label{DX1}
\end{equation}
where the $l$-independent functions $N_i^{\mu...}$ contain Dirac
structures and spinors from the numerator, $D[l]$ denotes
the denominator, which  behaves at large Euclidian momentum $l$ as
$D[l]\sim (l^2)^3$. Such situation is usual for the $UV-$divergent
graphs mentioned above, except only diagrams with quark
self-energy subgraphs. Substituting (\ref{DX1}) in (\ref{DX0}) we
obtain
\begin{equation}
D_{X}=N_2^{\mu\nu} J[l_\mu l_\nu]+N_1^{\mu}J[_\mu]+N_0 J_0,
\label{DX2}
\end{equation}
where we introduced
\begin{equation}
J[l_\mu l_\nu]=\int d^{D}l \frac{l_\mu l_\nu}{D[l]} \label{def:J}
\end{equation}
and similar for others integrals. Taking into account the behavior of
the denominator at large momentum $D[l]\sim (l^2)^3$ , it's clear that
only $J[l_\mu l_\nu]$ is $UV-$divergent.
 The other three integrals can have only $IR$-divergencies,
regulated by the off-shell momenta and therefore can be considered in
$D=4$. One can easily express the tensor integral
 $J[l_\mu l_\nu]$ through scalar integrals:
\begin{equation}
J[l_\mu l_\nu]= g_{\mu\nu}J_1+ (n_\mu \bar n_\nu + \bar n_\mu
n_\nu)J_2 \label{def:J12}
\end{equation}
with
\begin{eqnarray}
J_1& =&\frac{1}{2(1-\varepsilon)} \left( J[l^2]-J[(l\cdot
n)(l\cdot\bar n)]\right),
\label{def:J1}\\
J_2& =&\frac{1}{4(1-\varepsilon)} \left( (2-\varepsilon)J[(l\cdot
n)(l\cdot\bar n)]-J[l^2]\right).
\label{def:J2}
\end{eqnarray}
Both scalar integrals $J[l^2]$ and $J[(l\cdot n)(l\cdot\bar n)]$
have $UV-$poles. But in the coefficient $J_2$ the poles cancel.
Hence we must  contract $g_{\mu\nu}$ with Dirac
structure $N_2^{\mu\nu}$ in $D=4-2\varepsilon$ and expand the obtained
expression up to terms $\sim \varepsilon$. The reduction of all
one-loop Dirac structures to tree spinor combinations
can be  performed with the help of $NDR$ prescription%
\begin{equation}
\gamma^{\mu}\gamma^{\mu_{1}}\gamma^{\mu_{2}}(1-\gamma_{5})\otimes\gamma_{\mu
}\gamma_{\mu_{1}}\gamma_{\mu_{2}}(1-\gamma_{5})=4(4-\varepsilon)\gamma^{\mu
}(1-\gamma_{5})\otimes\gamma_{\mu}(1-\gamma_{5})
\end{equation}
Let us here make following observation. Computing the integrals
$J[l^2]$ and $J[(l\cdot n)(l\cdot\bar n)]$ we do not remove the
$IR-$regularization because these integrals
can also happen to be $IR-$divergent
and we must avoid the mixing of $UV-$ and $IR-$poles. However, we
observed, that one can always choose some``convenient''
 momentum flow for which these integrals have only $UV-$divergences and free from
$IR-$singularities. In this case one can drop the IR-regularization
making the calculations more simple.

Defining  $UV-$pole of the coefficient $J_1$ as
\begin{equation}
 {\rm pole\,\, part}\left[J_1\right] = \frac 1 \varepsilon Z^{UV}
\label{def:UVC1}
\end{equation}
we  write $UV-$divergent contribution for the $D_X$ as
\begin{equation}
 D^{UV}_X = \frac 1 \varepsilon Z^{UV} \lim_{\varepsilon\rightarrow0}
 N_2^{\mu\nu}g_{\mu\nu}
 =im^2_b f_P \xi_B \frac{\alpha_S}{2\pi}X_{col} JX_{UV}
\label{def:UVJ1}
\end{equation}
where $X_{col}$ is the color factor of the diagram and $JX_{UV}$
represents some $UV-$divergent expression. In  Appendix we provide
explicit expressions for $X_{col}$ and $JX_{UV}$ for each diagram.

The singular contributions $ D^{UV}_X$ are removed by tree diagrams
with operators renormalization constants and by QCD counterterms,
see details in the Appendix. After that, performing  Fierz
transformation, we obtaine trivial  contributions from all the
diagrams with topology $ A5-G5$ and $G\{1,3\}$.

The similar calculation of the self-energy one-loop diagrams is
much simple because they have only $UV-$poles and can be easily
reduced to the tree Dirac structures. Let us mention, that we must
consider also the diagrams with the one loop corrections to the
wave functions  which are not listed in the
Fig.\ref{Fig:NLOgraphs}, except for the heavy quark $\Sigma_b$. We
discuss these contributions in the Appendix.

Finally, the finite expression for the diagram under consideration  can be
written as
\begin{equation}
\left[D_{X}\right]_R=\left[ N_2^{\mu\nu}g_{\mu\nu} J_1 \right]_{R}
+N_2^{\mu\nu}(n_\mu \bar n_\nu + \bar n_\mu  n_\nu)J_2
+N_1^{\mu}J[l_\mu]+N_0 J_0, \label{DXren}
\end{equation}
where $[...]_R$ denotes renormalized quantity. Remaining
integrals have only $IR-$divergencies: collinear and soft.
Calculation of such contributions is the same for all diagrams,
with  and without $UV-$subgraphs.  Because we use off-shell
regularization, we put $D=4$ and
perform projections on the pseudoscalar state with momentum $p^{\prime}$ :%
\begin{equation}
u_{\beta}^{a}~\bar{u}_{\alpha}^{b}~=(-i)~f_{P}\frac{\delta^{ab}}{~N_{c}%
}~\left[  \slash {p}^{\prime}\gamma_{5}\right]  _{\alpha\beta},
\end{equation}
and on the \scetI operator $J^{B1}$:
\begin{align}
~\bar{d}^a_{\alpha}b^c_{v\beta}  &  =\bar{d}_n\, t^A
\gamma_{\perp}^{\sigma}h_{v}~\left[  V_{B1}^{\sigma}\right]
_{\beta\alpha
}\otimes t^A_{ca}+~...~,\\
\left[  V_{B1}^{\sigma}\right]  _{\beta\alpha}  &
=~\frac{1}{2}\left[
\Dirac{n}~\gamma_{\perp}^{\sigma}-\frac{1}{2}\Dirac{\bar
n}\gamma_{\perp}^{\sigma
}\Dirac{n}\right]  _{\beta\alpha}%
\label{Vproector}
\end{align}
where $t^A$ are the standard color matrices satisfying
Tr$(t^At^B)=\frac12\delta_{AB}$ and dots stand for the irrelevant spin
and color structures.  It is convenient also to insert
parametrisation for the vector integral
\begin{eqnarray}
J[l_\mu]=\frac 12 n_\mu J[(l\cdot \bar n)]+ \frac 12 \bar n_\mu
J[(l\cdot n)].
\end{eqnarray}
After calculation of traces and contractions we arrive to the
expression for $\left[D_{X}\right]_R$ which can be written as a sum scalar
integrals
\begin{align}
\left[D_{X}\right]_R  &  =
im_{b}^{2}~f_{P}\xi_{B}\frac{\alpha_{S}}{2\pi}~\left[J_{X}\right]_R,
\\
\left[J_{X}\right]_R& = \left[a_{21}(\varepsilon) J[l^2]+
a_{22}(\varepsilon) J[(l\cdot n)(l\cdot\bar n)]\right]_R + b_{11}
J[(l\cdot \bar n)]+ b_{12} J[(l\cdot n)]+ c_0 J_0\,\, .
\label{JXren}
\end{align}
with some coefficients $a_{2i},b_{1i}$ and $c_0$. The coefficients
$a_{2i}$, in front of $UV-$divergent integrals are
 computed in dimension $D$ and therefore depend on $\varepsilon$.

The sum of the
 integrals $\left[J_{X}\right]_R$ gives the formula
 for the form factor $\varkappa_{i}^{T}$:
\begin{align}
\left(  \varkappa_{i}^{T}~\right)  _{NLO}  &  =\sum
_{X}~\left[J_{X}\right]_R\, . \label{kappaJX}%
\end{align}
Let us stress again, that expression with brackets $[...]_R$ in
(\ref{JXren}) denotes the renormalized quantity. We simply have rewritten the
coefficients $J_{1,2}$ in terms of the corresponding integrals
(\ref{def:J2}),(\ref{def:J1}) and introduced
$\varepsilon$-dependent factors $a_{2i}$ which arise from the
calculations in DR. Assume for simplicity that $UV-$divergent
integrals are free from $IR-$singularities. As we have discussed
above, such situation can be realized for each diagram. Then
contributions associated with these integrals are simply some
finite expressions which do not depend on $IR-$regularization
parameters. Our task now is to compute the remaining integrals
$J[(l\cdot n)]$, $J[(l\cdot\bar n)]$ and $J_0$.

Evaluation of these four-dimensional integrals can be  performed
with the technique known as expansion by regions, see for instance
\cite{smirnov02}. The dominant regions have been discussed above
in the text. Hence, instead of one finite integral we obtain the
sum of more simple but divergent integrals. According to general
prescription, we use dimensional regularization in order to
regularize the simplified integrals in each region. Therefore, in
accordance with the dominant regions we have
find  following general decomposition of $\left[J_{X}\right]_R$%
\begin{equation}
\left[J_{X}\right]_R=\left(  J_{X}\right)  _{hard}+\left(  J_{X}\right)
_{coll-p}+\left( J_{X}\right)  _{coll-p^{\prime}}+\left(
J_{X}\right)  _{soft}~.
\label{Jregions}%
\end{equation}
Taking into account that $IR-$divergencies are related with
collinear and soft regions we find:
\begin{eqnarray}
\left(  J_{X}\right)  _{hard}&=&\left[a_{21}(\varepsilon) J[l^2]+
a_{22}(\varepsilon) J[(l\cdot n)(l\cdot\bar n)]\right]_R +
\nonumber \\ & &
\phantom{ddddddddddd}
\left( b_{11}J[(l\cdot \bar n)]+ b_{12} J[(l\cdot n)]+ c_0 J_0
\right)_{hard}\, ,
\label{def:Jhard}\\[3mm]
\left(  J_{X}\right)  _{coll,\, soft}&=&
\left( b_{11}J[(l\cdot \bar n)]+ b_{12} J[(l\cdot n)]+ c_0 J_0
\right)_{coll,\, soft}\, .
\label{def:collsoft}
\end{eqnarray}
 The hard region contributions in DR have $IR-$poles instead of $IR-$logarithms
 as in the case of off-shell regularization. The collinear and soft contributions have
 only $UV-$divergencies (off-shell regularization works in $IR-$regions).
 But the sum of all terms must be finite because of \ cancellation
 between $UV-$ and $IR-$poles.

 The contributions from the
collinear and soft regions depend on  the external off-shell
momenta\ which we use as $IR-$regulators in the original integral
$J_{X}$. Inserting decomposition (\ref{Jregions}) into
(\ref{kappaJX}) and then into (\ref{TNLO}) we must recover the
cancellation of the soft $IR-$scales.
 This is a good check of
the factorization in the next-to-leading order.
It is convenient to define the quantity:
\begin{eqnarray}
S&=&\sum_X\left\{
\left(  J_{X}\right)  _{coll-p}+\left(  J_{X}\right)  _{coll-p^{\prime}%
}+\left(  J_{X}\right)  _{soft}
\right\}
\nonumber \\[2mm] && \phantom{dddddddd}
+\frac{C_{i\pm1}}{N_{c}}
\left(  \int_{0}%
^{1}dx^{\prime}\frac{\varphi_{P}^{NLO}(x^{\prime})}{1-x^{\prime}}
+\frac{1}{\bar u}\int_{0}%
^{1}dz^{\prime}\theta^{NLO}(z^{\prime})\right)  ~.~
\label{def:s}
\end{eqnarray}
Then taking into account that $\varphi_{P}^{NLO}$ and $\theta^{NLO}$ are
defined as matrix elements of the renormalized operators ( the $UV-$poles are
subtracted )\footnote{ Note that one must apply the same $IR-$regularization
for the matrix elements which define \ $\varphi_{P}^{NLO}$ and $\theta^{NLO}%
$.} we expect that the answer for  $S$ can be represented as a sum of $UV-$poles and
scale independent constant:
\begin{equation}
S=\frac{1}{\varepsilon^{2}}Z_{2}+\frac{1}{\varepsilon}Z_{1}+Z_{0}.
\label{Seps}%
\end{equation}
The poles,  arising from the collinear and soft
integrals\ in (\ref{Seps}),
must cancel against  $IR-$poles
appearing in the hard integrals $\left(  J_{X}\right)  _{hard}$ in (\ref{Jregions}).
 It is clear that the
residues $Z_{1,2}$ can be related to the corresponding renormalization
constant of the LCDA $\varphi_{P}$
and \ form factor $\theta$.

In order to obtain the finite terms $Z_{0}$ one has to perform calculation of the
collinear and soft integrals, matrix elements (\ref{thetaPT1}) and
(\ref{phiP1}) and compute the sum. It is clear that both calculations
overlap and this
may be used to simplify derivation of the term $Z_{0}$. The \scetI non-renormilized
matrix elements
must be computed in DR, as usually, with $D$-dimensional Dirac algebra.
But the structure of numerators of the diagrams
for the matrix elements  are relatively
simple and
their reduction to the basic combinations can be evaluated without any
special prescriptions.
 The
calculation of the
form factor (\ref{thetaPT1}) is more complicate in comparison with
 pure collinear pion LCDA (\ref{phiP1}) because certain
diagrams generate $UV-$poles of second order.
But corresponding diagrams always
have very simple numerators.  Performing reduction of Dirac
algebra to the basic structures one can again rewrite
expressions  for
 the
matrix elements in terms of the scalar integrals which are  similar to
those  appearing from the collinear and soft regions in  QCD
diagrams. Now the coefficients in front of these integrals are
$\varepsilon$-dependent. We have found following representation
for the bare NLO \scetI matrix elements:
\begin{eqnarray}
&& \frac{C_{i\pm1}}{N_{c}}
\left(  \int_{0}%
^{1}dx^{\prime}\frac{\varphi_{P}^{NLO}(x^{\prime})}{1-x^{\prime}}
+\frac{1}{\bar u}\int_{0}%
^{1}dz^{\prime}\theta^{NLO}(z^{\prime})\right)_{bare}
=
\nonumber   \\[5mm]
& & \phantom{ddddddddddddd}
-\sum_X
\left\{
\left(  J_{X}\right)  _{coll-p}+\left(  J_{X}\right)  _{coll-p^{\prime}%
}+\left(  J_{X}\right)  _{soft}
\right\}
+\varepsilon\,\sum_X I_X\,
\label{NLObare}
\end{eqnarray}
where $I_X$ are some $UV-$divergent integrals.
These integrals have only
 first order poles in $\varepsilon$. All integrals with second order poles
 are absorbed into the first sum in rhs (\ref{NLObare}).
 Therefore taking into account that integrals $I_X$ have
 coefficient $\varepsilon$ we obtain finite contribution from the second term
 in rhs (\ref{NLObare}).
  The factor $\varepsilon$ in
the numerators appears, as a rule, from the reduction of $D$-dimensional structures
in the diagrams to the basic factors $\xi_B$ and $f_P$.
 Combining (\ref{NLObare}) with (\ref{def:s})
and taking into account $UV-$counteterterms for the \scetI operators we obtain
representation (\ref{Seps}).
It is clear that terms $ \varepsilon\, I_X\ $ provide contributions to the
constant $Z_0$ in (\ref{def:s}).

Substitution (\ref{ViNLO}) and (\ref{Seps}) in
formula for the coefficient function (\ref{TNLO}) gives:
\begin{equation}
T_{i}^{NLO}(u,z)=\sum_{X}\left(  J_{X}\right)  _{hard}+\frac{1}{\varepsilon
^{2}}Z_{2}+\frac{1}{\varepsilon}Z_{1}+Z_{0}+\frac{C_{_{i\pm1}}}{N_{c}}V(u).
\label{TNLO:1}%
\end{equation}
This is our final working formula.
It is convenient to rewrite the pole contributions as%
\begin{equation}
Z_{1}=(Z_{1})_{s/coll-p}+(Z_{1})_{coll-p^{\prime}}%
\end{equation}
where $(Z_{1})_{s/coll-p}$ \ \ and $(Z_{1})_{coll-p^{\prime}}$ \
 is associated with the \ form factor $\theta^{NLO}$ \ (soft and
collinear --$p$ regions) and LCDA $\ \varphi_{P}^{NLO}$ \
(collinear--$p^{\prime}$ region)\ \ respectively. Calculation of
the relevant integrals (details are considered in the Appendix) gives
\begin{equation}
(Z_{1})_{coll-p^{\prime}}=C_{F}\left(
-\frac{C_{_{i\pm1}}}{N_{c}}\right)  \frac{1}{\bar{u}}\left(
2+\ln\bar{u}\right), \label{Zet1}%
\end{equation}
\begin{align}
\frac{1}{\varepsilon^{2}}Z_{2}+\frac{1}{\varepsilon}(Z_{1})_{s/coll-p}
& = &  \left(  -\frac{C_{_{i\pm1}}}{N_{c}}\frac{1}{\bar{u}}\right)
\left(
-\frac{1}{\varepsilon^{2}}\frac{C_{F}}{2}(1+2\varepsilon\ln\left[
\mu /(p\bar n)\right]  )  +\frac{1}{\varepsilon}\left[
C_{F}\frac{\ln
z}{\bar z}-\frac{C_{A}}{2}\frac{\bar z+\ln z}{1-z}\right]  \right)  \label{Zet21}%
\end{align}
We checked that above expressions are consistent with the evolution
kernels of the pion LCDA \cite{RE78,BL80} and  \scetI operator $J^{B1}$
\cite{BY05,NBH}.
But  note, that we haven't included the diagrams with the wave function
 renormalizations hence the poles
(\ref{Zet1}) and (\ref{Zet21}) are not
exactly convolutions of the corresponding evolution kernels with the leading order
coefficient functions.

Calculation of the hard integrals $\left(  J_{X}\right)  _{hard}$
can be done with standard technique and the results for each
diagram are listed in the Appendix. The arising $IR-$poles cancel in the
sum with (\ref{Zet1}) and (\ref{Zet21}) as it is expected from the
factorization theorem. Resulting expressions for the coefficient
functions can be written as%
\begin{align}
\alpha_{2}  &  :~T_{2}^{NLO}(u,z)=\frac{C_{2}}{2N_{c}}T_{D}(u,z)+\frac{C_{1}%
}{N_{c}}T_{ND}(u,z),\\
\alpha_{1}  &  :~T_{1}^{NLO}(u,z)=\frac{C_{1}}{2N_{c}}T_{D}(u,z)+\frac{C_{2}%
}{N_{c}}T_{ND}(u,z),
\end{align}
where the subscripts $D~$\ and $ND$ can be understood as diagonal and non-diagonal
contributions. Assuming for simplicity $\mu_R=\mu_F=\mu$ we obtaine:
\begin{align}
\frac{1}{\pi}\operatorname{Im}T_{D}(u,z)  &  =\frac{1}{2}\left(  \frac
{4-u}{\bar{u}}-\frac{2\,u^{2}}{{\left(  1-u-z\right)  }^{2}}-\frac{u}%
{1-u-z}\right) \nonumber\\
&  +\left(  \frac{1}{\bar{u}}-\frac{\,\bar{z}~u^{2}}{{\left(  1-u-z\right)
}^{3}}\right)  \,\ln u\nonumber\\
&  +\frac{z\,}{u-z}\ln\bar{u}+u\,\left(  -\frac{1}{\bar{u}}-\frac{1}{u-z}%
+\frac{u~\,\bar{z}}{{\left(  1-u-z\right)  }^{3}}\right)  \,\ln\bar{z}~,
\label{ImT22}%
\end{align}%
\begin{align}
\operatorname{Re}~T_{D}(u,z)  &  =\frac{3}{\bar{u}}\ln\left(  \mu^{2}%
/m_{b}^{2}\right)  +\frac{8}{\bar{u}}+\frac{1}{2}\left(  -\frac{u^{2}%
+3\,\bar{u}}{{\bar{u}}^{2}}+\frac{2\,u^{2}}{{\left(  1-u-z\right)  }^{2}%
}+\frac{u}{1-u-z}-\frac{u}{{\bar{u}}^{2}\,\bar{z}}\right)  \,\ln u\nonumber\\
&  \,+\left(  \frac{1}{2\,{\bar{u}}^{2}\,\bar{z}}-\frac{u}{\bar{u}\,{\left(
1-u-z\right)  }^{2}}-\frac{u}{2\,{\bar{u}}^{2}\,\left(  1-u-z\right)
}\right)  \,\left(  1-\bar{u}\,z\right)  \ln(1-\bar{u}\,z)\nonumber\\
&  +\frac{\left(  1-u\,z\right)  \,}{u\,\bar{u}\,z\,\bar{z}}\ln
(1-u\,z)-\left(  \frac{3}{2\,\bar{u}}+\frac{1}{u\,\bar{z}}\right)  \,\ln
\bar{u}-\frac{\left(  2-3\,z\right)  }{2\,\bar{u}\,\bar{z}}\,\ln z\nonumber\\
&  -\frac{\bar{z}\,\left(  \bar{z}-3\,u\right)  \,}{2\,{\left(  1-u-z\right)
}^{2}}\ln\bar{z}-\frac{{\ln}^{2}{\bar{u}}}{2\,\bar{u}}+\frac{{\ln}^{2}{u}%
}{2\,\bar{u}}+\frac{\,\ln z}{\bar{u}}\left(  \ln\bar{u}-\ln u\right)
\nonumber\\
&  +\frac{\mathrm{Li}_{2}(\bar{u})-\mathrm{Li}_{2}(u)}{\bar{u}}-\frac{\bar
{z}\,~u^{2}\,}{{\left(  1-u-z\right)  }^{2}}I(u,\overline{z})+\frac{u\,\bar
{z}\,}{\bar{u}}I(\bar{u},\overline{z})~, \label{ReT22}%
\end{align}%
\begin{align}
\frac{1}{\pi}\operatorname{Im}T_{ND}(u,z)  &  =\frac{C_{A}}{2}\,\left(
\frac{u}{1-u-z}-\frac{1+2\,u}{2\,\bar{u}}\right. \nonumber\\
&  -\frac{z\,}{u-z}\ln\bar{u}+\left(  \frac{u^{2}}{{\left(  1-u-z\right)  }%
^{2}}-1\right)  \,\ln u\nonumber\\
&  \left.  +\frac{z\,}{\bar{u}\,\bar{z}}\ln z+\left(  \frac{1}{\bar{u}%
}\,-\frac{u^{2}}{{\left(  1-u-z\right)  }^{2}}+\frac{u}{u-z}\right)  \,\ln
\bar{z}\right)  \nonumber\\
&  +{C_{F}}\,\left(  \frac{1}{2}\left(  \frac{5-u}{\bar{u}}-\frac{2\,u^{2}%
}{{\left(  1-u-z\right)  }^{2}}-\frac{u}{1-u-z}\right)  \right. \nonumber\\
&  +\left(  \frac{1}{\bar{u}}-\frac{u^{2}\,\bar{z}}{{\left(  1-u-z\right)
}^{3}}\right)  \,\ln u+\frac{u\,\bar{z}\,}{\bar{u}\,\left(  u-z\right)  }%
\ln\bar{u}-\frac{z\,}{\bar{u}\,\bar{z}}\ln z\nonumber\\
&  +\left.  \,\left(  \frac{u^{2}\,\bar{z}}{{\left(  1-u-z\right)  }^{3}}%
-\frac{u}{\bar{u}}-\frac{u}{u-z}\right)  \,\ln\bar{z}\right)  \label{ImT21}%
\end{align}

\begin{equation}
\operatorname{Re}~T_{ND}(u,z)={C_{F}~T_{CF}(u,z)+}\frac{C_{A}}{2}%
~T{_{CA}(u,z),}%
\end{equation}%

\begin{align}
T{_{CF}}(u,z)  &  =\frac{1}{4\bar{u}}\ln^{2}\left(  \mu^{2}/m_{b}^{2}\right)
+\frac{1}{\bar{u}}\left(  \frac{21}{4}+\ln\bar{u}-\frac{\ln z}{\bar{z}%
}\right)  \ln\left(  \mu^{2}/m_{b}^{2}\right)
\nonumber
\end{align}%
$$
 +\frac{27}{2\,\bar{u}}+\frac{{\pi}^{2}\,\left(  4-3\,u-5\,u\,z-4\,u^{2}%
\,\bar{z}\right)  }{24\,u\,\bar{u}\,\bar{z}}-\left(  \frac{1}{2\,\bar{u}%
}+\frac{2}{u\,\bar{z}}\right)  \,\ln\bar{u}
$$
$$  +\left(  -\frac{2+u}{\bar{u}}+\frac{u^{2}}{{\left(  1-u-z\right)  }^{2}%
}+\frac{u}{2\,\left(  1-u-z\right)
}+\frac{1}{\bar{u}\,\bar{z}}\right)  \,\ln u
$$
%
$$  -\frac{3\,\left(  3-2\,z-u\,\bar{z}\right)  \,}{2\,\bar{u}\,\bar{z}}\ln
z+\frac{2\,\left(  1-u\,z\right)  \,}{u\,\bar{u}\,z\,\bar{z}}\ln(1-u\,z)
$$
$$
  +\frac{1}{2}\left(  -1-\frac{3}{\bar{u}}+\frac{2\,u^{2}}{{\left(
1-u-z\right)  }^{2}}+\frac{u}{1-u-z}-\frac{1}{u\,z^{2}}+\frac{3-5\,u}%
{u\,\bar{u}\,z}\right)  \,\ln\bar{z}
$$
$$  +\left(  \frac{1}{2\,u\,\bar{u}\,z^{2}}-1-\frac{\left(  2-u\right)  \,u^{2}%
}{\bar{u}\,{\left(  1-u-z\right)  }^{2}}-\frac{u\,\left(  u^{2}+2\,\bar
{u}\right)  }{2\,{\bar{u}}^{2}\,\left(  1-u-z\right)  }\right.
$$
$$  +\left.  \frac{6\,u-3-4\,u^{2}}{2\,u\,{\bar{u}}^{2}\,z}-\frac{1}{\bar
{u}\,\bar{z}}\right)  \ln(1-\bar{u}\,z)
$$ 
$$  +\frac{\ln\bar{u}\,}{\bar{u}}\left(  -\left(  1+\bar{u}\right)  \,\ln
u+\frac{1+\bar{u}\,\bar{z}}{\bar{z}}\ln z-\ln\bar{z}\right)
$$ 
$$ -\frac{{\ln}^{2}{\bar{u}}}{\bar{u}}+\frac{{\ln}^{2}{u}}{2\,\bar{u}}%
+\frac{\left(  2-z\right)  }{2\,\bar{u}\,\bar{z}}\ln^{2}{z}-\frac{\left(
1+\bar{u}\right)  \,}{\bar{u}}\ln u\,\ln z
$$
$$+\frac{z-2\,\bar{u}\,\bar{z}\,}{\bar{u}\,\bar{z}}\mathrm{Li_{2}}(u)-\frac
{1}{u\,\bar{u}~\,\bar{z}}\mathrm{Li_{2}}(\bar{u})-\frac{1-u\,\left(
3-z\right)  }{u\,\bar{u}~\,\bar{z}}\,\mathrm{Li_{2}}(z)
$$ 
$$-\frac{\left(  1+u\,\bar{z}\right)  }{\bar{u}\,\bar{z}}\,\mathrm{Li_{2}%
}(u\,z)+\left(  \frac{1}{u\,\bar{u}\,\bar{z}}-1\right)  \,\mathrm{Li_{2}}%
(\bar{u}\,z)+\frac{1}{\bar{u}}\mathrm{Li_{2}}(\bar{z})
$$ 
\begin{align}
-\frac{u^{2}\,\bar{z}\,}{{\left(  1-u-z\right)  }^{2}}I(u,\overline
{z})+\frac{u\,\bar{z}\,}{\bar{u}}I(\bar{u},\overline{z})~,
\label{TCF}%
\end{align}

\begin{align}
T{_{CA}}(u,z)  &  =-\frac{1}{\bar{u}}\left(  3-\frac{\ln z}{\bar{z}}\right)
\ln\left(  \mu^{2}/m_{b}^{2}\right)  -\frac{9}{\bar{u}}-\frac{{\pi}%
^{2}\,\left(  1-u\,\left(  3-z\right)  \right)  }{6\,u\,\bar{u}\,\bar{z}%
}\nonumber\\
\end{align}
$$
+\frac{\ln\bar{u}}{{u}\,\bar{z}}+\frac{3\,\left(  2-z\right)  \,}{2\,\bar
{u}\,\bar{z}}\ln z-\left(  1-\frac{5}{2\,\bar{u}}+\frac{u}{1-u-z}+\frac
{1}{\bar{u}\,\bar{z}}\right)  \,\ln u
$$
$$-\frac{1-u\,z}{u\,\bar{u}\,z\,\bar{z}}\,\ln(1-u\,z)+\frac{\left(  1-\bar
{u}\,z\right)  \,\left(  \bar{z}-u\,\left(  1-z\,\bar{z}\right)  \right)
\,}{u\,\bar{u}\,\left(  1-u-z\right)  \,z\,\bar{z}}\ln(1-\bar{u}%
\,z)
$$ 
$$ -\left(  1-\frac{3}{2\,\overline{u}}+\frac{u}{1-u-z}+\frac{1-2\,u}%
{u\,\bar{u}\,z}\right)  \,\ln\bar{z}
$$ 
$$-\frac{1}{2\,\bar{u}}\,\left(  1+\frac{1}{\bar{z}}\right)  {\ln}^{2}%
{z}+\frac{{\ln}^{2}{\bar{u}}}{2}-\frac{{\ln}^{2}{u}}{2}+\frac{1}{2\,\bar{u}%
}{\ln}^{2}{\bar{z}}
$$
$$-\ln\bar{u}\,\left(  \frac{1}{\bar{u}\,\bar{z}}\ln z+\ln\bar{z}\right)
+\frac{\ln u}{\bar{u}}\,\left(  \ln z-{u}\,\ln\bar{z}\right)
$$
$$+\left(  1-\frac{1+u\,\bar{z}}{\bar{u}\,\bar{z}}\right)  \,\mathrm{Li_{2}%
}(u)-\left(  2-\frac{1}{u\,\bar{u}\,\bar{z}}\right)  \,\mathrm{Li_{2}}(\bar
{u})
$$ 
$$ +\frac{1}{\bar{u}}\mathrm{Li_{2}}(z)+\frac{\left(  1-3u+u~z\right)  }%
{{u}\,\bar{u}\,\bar{z}}\,\mathrm{Li_{2}}(z)+\frac{\left(  1+u\,\bar{z}\right)
\,}{\bar{u}\,\bar{z}}\mathrm{Li_{2}}(u\,z)
$$
\begin{align}
+ \left(  1-\frac{1}{u\,\bar{u}\,\bar{z}}\right)  \,\mathrm{Li_{2}}(\bar
{u}\,z)-\frac{1}{\bar{u}}\mathrm{Li_{2}}(\bar{z})+\frac{u^{2}}{1-u-z}%
\,I(u,\overline{z})-u\,~I(\bar{u},\overline{z})~. \label{TCA}%
\end{align}
 where we introduced convenient real function  $I(u,z)$ which reads%
\begin{align}
I(u,z)  &  =\frac{\ln(u/z)}{u-z}\left[  \frac{1}{2}\ln(uz)-\ln(\bar
{z}\,\bar {u})-\ln(u+z)\right]  +\frac{\ln(\bar {u}/\bar {z})}%
{u-z}  \ln\left[\frac{u+z}{uz}-1\right]  \nonumber\\
& + \frac{1}{u-z}\left[  2\mathrm{Li}_{2}(\bar {z})+\mathrm{Li}_{2}\left(
\frac{z}{u+z}\right)  +\mathrm{Li}_{2}\left(  \frac{z^{2}}{z+u\bar z}\right)
-(z\leftrightarrow u)\right] . \label{IC3}%
\end{align}

Our results are in agreement with the kernels which have been
earlier presented in the paper \cite{Beneke05} but computed using
different technical approach ( dimensional regularization with
evanescent operators)\footnote{ We are grateful to S.~J\"ager and
M.~Beneke for the correspondence which helped us to fix a mistake
in the expression for one diagram}.
 Let us also observe that the
function $I(u,z)$ introduced in eq.(\ref{IC3}) naturally appears
in calculations of the diagrams involving massive propagator of
heavy quark. The similar structure have been also introduced in
\cite{Beneke05} and denoted as $F(z,u)$. Let us stress once more, that
$I(u,z)$ is a real function as one can easily see from its
definition (\ref{IC3}). We have found that
\begin{equation}
I(u,z)= \frac1{z-u}{\rm Re}[F(1-z,1-u)].
\label{IvF}
\end{equation}
 We did not find any transformation to prove
this equivalence analytically and
 checked it numerically for the several arbitrary values of the arguments.

Analytical expressions (\ref{ImT22})-(\ref{TCA}) for the coefficient
 functions $T^{NLO}_{1,2}$
represent the main technical results of our paper.

\section{\textbf{Numerical estimates of
$B\rightarrow \pi\pi$ branching fractions}}

In this section we perform the numerical analysis of the branching fractions
including next-to-leading corrections to the hard and jet
coefficient functions.
The main contribution to the branchings originate, obviously,  from the
real parts of amplitudes $\alpha_{1,2}$. We neglect in our
estimates by electroweak penguins
$\alpha_{i,EW}$ but include QCD penguins $\alpha^{u,c}_4$, see eq.(\ref{Apipi}),
 in the form presented in \cite{BBNS}.

 Consider first
 some important details in calculation  $\alpha_{1,2}$ at the NLO
approximation. General formula reads:
\begin{align}
\alpha_{i}  &  =f_{0}\int_{0}^{1}du~V_{i}(u,\mu_{R},\mu_h)\varphi_{\pi}(u,\mu
_{h})+\\
&  ~~~~~~~~\int_{0}^{1}du~\varphi_{\pi}(u,\mu_{h})\int_{0}^{1}dz~T_{i}%
(u,z,\mu_{R},\mu_{h})~\xi_{\pi}^{B}(z,\mu_{h})~,\\
\xi_{\pi}^{B}(z,\mu_{h})  &  =\hat{f}_{B}f_{\pi}\int_{0}^{\infty}d\omega
\int_{0}^{1}dx~~\phi_{B}(\omega,\mu_{F})J(z,x,\omega,\mu_{h},\mu
_{F})~\varphi_{\pi}(x,\mu_{F}).
\end{align}
where we have shown explicitly the scale dependence. To estimate the values of
these amplitudes we use the following numerical input. For the coefficient
functions $C_{i=1,2}(\mu_{R})$ in the effective Hamiltonian (\ref{Heff}) we
employ the NLO results at $\ \mu_{R}=m_{b}~$ obtained in \cite{Buras}
($NDR$-scheme, NLO approximation):
\begin{equation}
~C_{1}^{NLO}(m_{b})=1.075,~~C_{2}^{NLO}(m_{b})=-0.170.
\end{equation}
where $m_{b}$ denotes  $b$-meson pole mass $m_{b}=4.8~$GeV. For two others
scales we accept the following values:%
\begin{equation}
\mu_{h}=m_{b},~~\mu_{F}=\mu_{hc}=1.5~\text{GeV}.
\end{equation}
\ For simplicity, the uncertainties in the scales setting \ will be ignored
in our analysis.
Then we obtained \ \
\begin{equation}
\alpha_{i}=f_{0}\int_{0}^{1}du~V_{i}(u,m_{b})\varphi_{\pi}(u,m_{b})+\int
_{0}^{1}du~\varphi_{\pi}(u,m_{b})\int_{0}^{1}dz~T_{i}(u,z,m_{b})~\xi_{\pi}%
^{B}(z,m_{b})~,
\end{equation}
To compute the form factor $~\xi_{\pi}^{B1}(z,m_{b})$ we must perform  evolution
from scale $m_{b}$ to the scale $\mu_{hc}$:%
\begin{equation}
~\xi_{\pi}^{B}(z,m_{b})~=\int_{0}^{1}dz^{\prime}U_{B}(z,z^{\prime},m_{b}%
,\mu_{hc})~\xi_{\pi}^{B}(z^{\prime},\mu_{hc}),
\end{equation}
where the evolution operator $U_{B}$ is derived from the solution of the
evolution equation \cite{BY05,NBH}:%
\begin{equation}
\frac{d}{d\ln\mu}~\xi_{\pi}^{B}(z,\mu)=-\int_{0}^{1}dz~[V(z,z^{\prime}%
)-\delta(z-z^{\prime})\Gamma_{cusp}(\alpha_{S})\ln\frac{\mu}{m_{b}}]\xi_{\pi
}^{B}(z^{\prime},\mu),~
\end{equation}
The explicit expression for the evolution kernels in our notation
are the same as in \cite{NBH},
see Eq's (46,47). Let us write the solution for $U_{B}$ in the form%
\begin{equation}
U_{B}=U_{B}^{LL}(z,z^{\prime})+U_{B}^{NLL}(z,z^{\prime})+..., \label{UB}%
\end{equation}
where $U_{B}^{LL}$ and $U_{B}^{NLL}$ can be understood as "leading
log" and "next-to-leading log" approximations. Both terms are
needed to perform complete calculation at  next-to-leading
order. The evolution of the form factor is combination of the two
effects: summation of the so-called Sudakov double logarithms
associated with $cusp$-anomalous dimension $\Gamma_{cusp}$ and
usual light-cone logarithms associated with non-local evolution
kernel $V(z,z^{\prime})$ (or non-local anomalous dimension). At
present $\Gamma _{cusp}(\alpha_{S})$ and $V(z,z^{\prime})$ are
known at two- \ and one-loop accuracy respectively. This is enough
for summation of leading logarithms in $U_{B}^{LL}$ but in order
to compute $U_{B}^{NLL}$ one has to know $\Gamma_{cusp}~$ at three
loop and $V(z,z^{\prime})$ at two loop accuracy . Because these
quantities are uncalculated, we can't perform corresponding
evolution. So we just neglect by this effect in our calculation.

 Form factor $\xi_{\pi}^{B}(z,\mu_{hc})$ can be computed systematically
order by order using factorization approach . Corresponding jet function has
been computed at the next-to-leading accuracy in \cite{BHjet,BY05,Krn}. Fixing the
factorization scale at this step to be equal to \ $\mu_{F}=\mu_{hc}$ we escape
the summation of the large logarithms in jet function and therefore we have to
provide the  LCDA's $\varphi_{\pi}(x,\mu_{hc})$ and $\phi
_{B}(\omega,\mu_{hc})$ at this normalization point. Assuming decomposition
 $\ \xi_{\pi}%
^{B}=(\xi_{\pi}^{B})^{LO}+\frac{\alpha_{S}}{2\pi}(\xiB)^{NLO}$ we obtaine:
\begin{align}
\alpha_{i}^{HSA}  &  =\int_{0}^{1}du~\varphi_{\pi}(u,m_{b})~\int_{0}%
^{1}dz~T_{i}(u,z)U_{B}(z,z^{\prime})\ast\left[  (\xiB)^{LO}(z^{\prime
})+\frac{\alpha_{S}}{2\pi}(\xiB)^{NLO}(z^{\prime})\right]
\\
&  =\int_{0}^{1}du~\varphi_{\pi}(u)~T_{i}^{LO}(u)\int_{0}^{1}dz~~U_{B}%
^{LL}(z,z^{\prime})\ast(\xiB)^{LO}(z')+
 \nonumber \\
&  \int_{0}^{1}du~\varphi_{\pi}(u)~T_{i}^{LO}(u)\int_{0}^{1}dz~U_{B}%
^{LL}(z,z^{\prime})\ast\left[  \frac{\alpha_{S}}{2\pi}(\xiB)^{NLO}%
(z')\right]  +
\label{alpha i HSA}
\\
&  \int_{0}^{1}du~\varphi_{\pi}(u)\int_{0}^{1}dz\left[  ~\frac{\alpha_{S}%
}{2\pi}T_{i}^{NLO}(u,z)\right]  U_{B}^{LL}(z,z^{\prime})\ast(\xiB)%
^{LO}(z^{\prime})+...
\nonumber
\end{align}
where dots denote the neglected logarithms associated with $\ U_{B}^{NLL}$ and
higher order terms $O(\alpha_{S}^{2})$. For simplicity by asterisk we denoted
the integration with respect to $z^{\prime}$ and skip obvious scale dependence.
\ In the first two lines of $\left(  \ref{alpha i HSA}\right)  $we used that
$T_{i}^{LO}$ does not depend on fraction $z$. Then one can perform
integration  over external variable $z$%
\begin{equation}
\int_{0}^{1}dz~~U_{B}^{LL}(z,z^{\prime})=U^{LL}(z^{\prime})
\end{equation}
that simplifies the evolution. In the numerical calculations we have used for
$U^{LL}(z^{\prime})$ simple approximation that was found in \cite{BY05}.
Therefore we obtained
\begin{equation}
\int_{0}^{1}dz~~U_{B}^{LL}(z,z^{\prime})\ast(\xiB)^{BLO}(z^{\prime
})=~\left(  -\pi\alpha_{S}(\mu_{hc})~\frac{C_{F}}{N_{c}}\frac{f_{B}f_{\pi}%
}{~~K_{F}~\lambda_{B}~m_{b}}\right)  \int_{0}^{1}\frac{dz~}{\bar{z}}%
~\varphi_{\pi}(z,\mu_{hc})U^{LL}(z),
\end{equation}
where we used standard notation%
\begin{equation}
~\lambda_{B}^{-1}=\int\frac{d\omega}{\omega}\phi_{B}(\omega,\mu_{hc})~.
\end{equation}
Assuming the following ansatz for the pion DA amplitude %
\begin{equation}
\varphi_{\pi}(u,\mu_{hc})=6u(1-u)+a_{2}^{\pi}(\mu_{hc})C_{2}^{3/2}%
(2u-1)+a_{4}^{\pi}(\mu_{hc})C_{4}^{3/2}(2u-1), \label{pionDAmod}%
\end{equation}
one obtains (useful technical details can be fond in \cite{BY05} ):%
\begin{equation}
\int_{0}^{1}\frac{dz~}{\bar{z}}~\varphi_{\pi}(z)U^{LL}(z)=2.72~(1+~a_{2}^{\pi
}+~a_{4}^{\pi}).
\end{equation}
The similar calculation for the next-to-leading contribution gives%
\begin{equation}
\int_{0}^{1}dz~U_{B}^{LL}(z,z^{\prime})\ast\left[  \frac{\alpha_{S}}{2\pi}%
(\xiB)^{NLO}(z')\right]
=\int dz'~U^{LL}(z')\frac{\alpha_{S}}{2\pi}(\xiB)^{NLO}(z',\mu_{hc})%
\end{equation}
\begin{equation}
=\alpha_{S}\,f_{B}f_{\pi}/K_{F}\int_{0}^{1}dz'\,U^{LL}(z')\int_{0}^{\infty}d\omega\int
_{0}^{1}dx~~\phi_{B}(\omega)~\left[  \frac{\alpha_{S}}{2\pi}%
J_{NLO}(z',x,\omega)\right]  ~\varphi_{\pi}(x)
\end{equation}%
\begin{align}
&  =\left(  -\pi\alpha_{S}~\frac{C_{F}}{N_{c}}\frac{f_{B}f_{\pi}}%
{~K_{F}~\lambda_{B}~m_{b}}\right)  \frac{\alpha_{S}}{\pi}~\left(  ~\left(
1+a_{2}^{\pi}+a_{4}^{\pi}\right)  \left\langle L^{2}\right\rangle -\left[
3.93+8.15~a_{2}^{\pi}+~10.05~a_{4}^{\pi}\right]  \left\langle L\right\rangle
+\right. \\
&  \left.  \left[  3.0+10.10a_{2}^{\pi}+16.60~a_{4}^{\pi}\right]  ~\right)  ,
\end{align}
where we introduced%
\begin{equation}
\left\langle L\right\rangle =\lambda_{B}\int_{0}^{\infty}d\omega~\phi
_{B}(\omega)\ln\left[  \frac{m_{b}\omega}{\mu_{hc}^{2}}\right]
,~~~\left\langle L^{2}\right\rangle =\lambda_{B}\int_{0}^{\infty}d\omega
~\phi_{B}(\omega)\ln^{2}\left[  \frac{m_{b}\omega}{\mu_{hc}^{2}}\right]  .
\label{phiBmom}%
\end{equation}
The last term in eq.$\left(  \ref{alpha i HSA}\right)  $ with the convolution
of the next-to-leading coefficient function can be represented
in following form:%
\begin{equation}
\int_{0}^{1}du~\varphi_{\pi}(u)\int_{0}^{1}dz~\frac{\alpha_{S}}{2\pi}%
T_{i}^{NLO}(u,z)\int_{0}^{1}dz^{\prime}U_{B}^{LL}(z,z^{\prime})(\xiB)%
^{LO}(z^{\prime},\mu_{hc})~~=
\end{equation}%
\begin{equation}
=~\left(  -\pi\alpha_{S}(\mu_{hc})~\frac{C_{F}}{N_{c}}\frac{f_{B}f_{\pi}%
}{K_{F}~~\lambda_{B}~m_{b}}\right)  \frac{\alpha_{S}}{2\pi}\left[
\sum_{m,n=0,2,4}t_{i}^{~mn}a_{\pi}^{m}(m_{b})a_{\pi}^{n}(\mu_{hc})\right]
\end{equation}
where the moments $t_{i}^{~mn}:$%

\begin{align}
t_{i}^{mn}  &  =\frac{C_{i}}{2N_{c}}t_{D}^{mn}+\frac{C_{i\pm1}}{N_{c}}%
t_{ND}^{mn},~~\\
t_{D(ND)}^{mn}  &  =\int_{0}^{1}du~6u\bar{u}~C_{m}^{3/2}(u-\bar{u})\int
_{0}^{1}dz~T_{D(ND)}^{NLO}(u,z)\int_{0}^{1}dz^{\prime}U_{B}^{LL}(z,z^{\prime
})6z~C_{n}^{3/2}(z-\bar{z})
\end{align}
have been computed numerically. If the evolution is switched off  our results
are in agreement with the moments computed in \cite{Beneke05},
see equations (54,55). Performing the evolution and with the scale
fixed as described above we obtained:%
\begin{equation}%
\begin{array}
[c]{ccc}%
t_{D}^{00}=73.64+13.85i\pi & t_{D}^{02}=66.80+16.70i\pi & t_{D}^{04}%
=66.85+17.74i\pi\\
t_{D}^{20}=27.94+21.22i\pi & t_{D}^{22}=23.86+28.46i\pi & t_{D}^{24}%
=24.44+32.10i\pi\\
t_{D}^{40}=-12.74+22.61i\pi & t_{D}^{42}=-9.11+31.37i\pi & t_{D}%
^{44}=-7.71+34.67i\pi
\end{array}
\end{equation}%
\begin{equation}%
\begin{array}
[c]{ccc}%
t_{ND}^{00}=-7.34-14.24i\pi & t_{ND}^{02}=33.15-17.43i\pi & t_{ND}%
^{04}=69.63-18.21i\pi\\
t_{ND}^{20}=-201.69-39.87i\pi & t_{ND}^{22}=-180.99-39.19i\pi & t_{ND}%
^{24}=-164.74-39.28i\pi\\
t_{ND}^{40}=-371.10-51.01i\pi & t_{ND}^{42}=-347.00-49.66i\pi & t_{ND}%
^{44}=-335.62-49.63i\pi
\end{array}
\end{equation}

In order to compute branching fractions we used the set of input parameters
given in the table below. The values for the coefficient functions $C_{3,4,6,8eff}$
are taken also from \cite{Buras}.\\[3mm]
\centerline{Table 1. Numerical values of the phenomenological parameters used to
 compute branching fractions.}
\begin{center}
\begin{tabular}
[c]{|l|l|}\hline
& {}\\[-.3cm]
$f_{\pi}$ & 131 MeV
 \\[0.1cm] \hline  & {}\\[-.3cm]
$a_{2}^{\pi}(1.5$GeV$)$ & 0.25$\pm0.15$
 \\[0.1cm] \hline  & {}\\[-.3cm]
$a_{4}^{\pi}(1.5$GeV$)$ & -0.05$\pm0.15$%
\\[0.1cm]\hline
\end{tabular}
\begin{tabular}
[c]{|l|l|}\hline
& {}\\[-.3cm]
$f_{B}$ & 200$\pm30~$MeV
 \\[0.1cm] \hline  & {}\\[-.3cm]
$\lambda_{B}(1.5$GeV$)$ & 0.35$\pm.15~$GeV
\\[0.1cm] \hline
& {}\\[-.3cm]
$f_{0}$ & 0.28$\pm0.05$%
\\[0.1cm]\hline
\end{tabular}
\
\begin{tabular}
[c]{|l|l|}%
\hline
& {}\\[-.3cm]
$m_{B}$ & 5.28GeV
 \\[0.1cm] \hline  & {}\\[-.3cm]
$m_{b}^{pole}$ & 4.8GeV
 \\[0.1cm] \hline  & {}\\[-.3cm]
$m_{c}^{pole}$ & 1.8GeV
\\[0.1cm]\hline
\end{tabular}
\\
\begin{tabular}
[c]{|c|c|c|c|c|c|c|c|}%
\hline {} &{} &{} &{} &{} &{} & {}&
\\[-.3cm]
$\left\vert V_{ub}\right\vert \times10^{3}$ & $\left\vert V_{cb}\right\vert
\times10^{3}$ & $\gamma$ & $\tau_{B},~$ps & $C_{3}(m_{b})$ & $C_{4}(m_{b})$ &
$C_{6}(m_{b})$ & $C_{8eff}(m_{b})$
\\[0.1cm] \hline  %
 {} &{} &{} &{} &{} &{} & {}&
\\[-.3cm]
$3.7^{+1.3}_{-0.8}$\cite{BABARf0} & 41.4$$ & 62$%
{{}^\circ}%
$ & 1.53 & 0.012 & -0.030 & -0.035 & -0.143
\\[0.1cm] \hline
\end{tabular}
\end{center}

For the pion LCDA we use a simple model with two Gegenbauer moments $a_{2,4}^{\pi}$
(\ref{pionDAmod}). Our estimates of the moments based on the results obtained in
\cite{Ball:2006wn,Bakulev:2005cp}.
The evolution of these moments from initial scale
 $\mu_{hc}$ to scale $m_b$ have been
computed with the next-to-leading logarithmic accuracy  for the leading
order contribution
(the first term in (\ref{alpha i HSA})).
To perform  this two-loop evolution we have used the analytical results
derived in \cite{DMueller}.

\begin{figure}[ptb]
\begin{center}
{\includegraphics[
height=1.4079in,
width=2.2675in
]%
{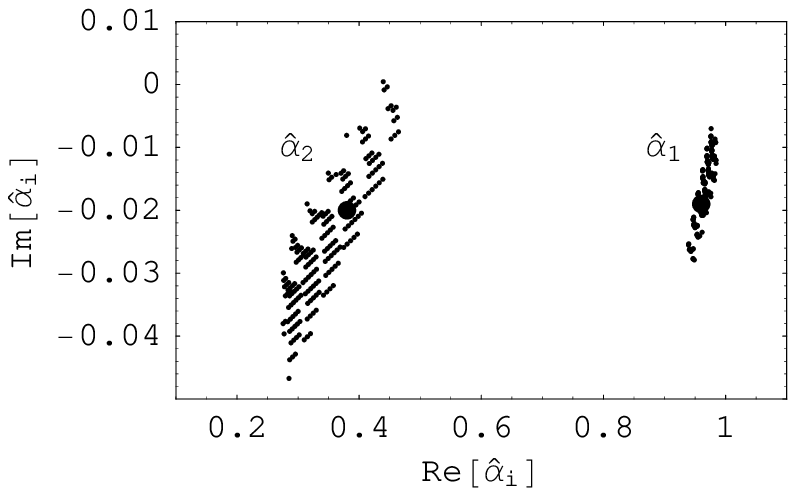}%
}%
\phantom{asdfghj}
{\includegraphics[
height=1.3612in,
width=2.1906in
]%
{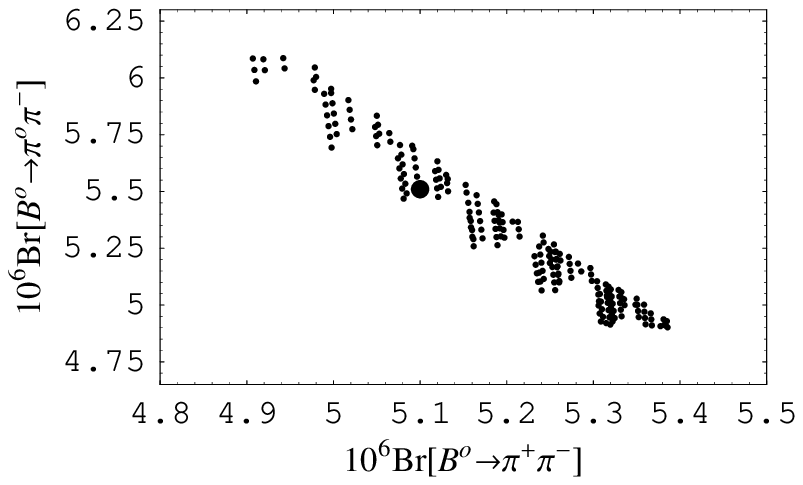}%
}
\\[2mm]
{\includegraphics[ height=1.4079in, width=2.2675in
]%
{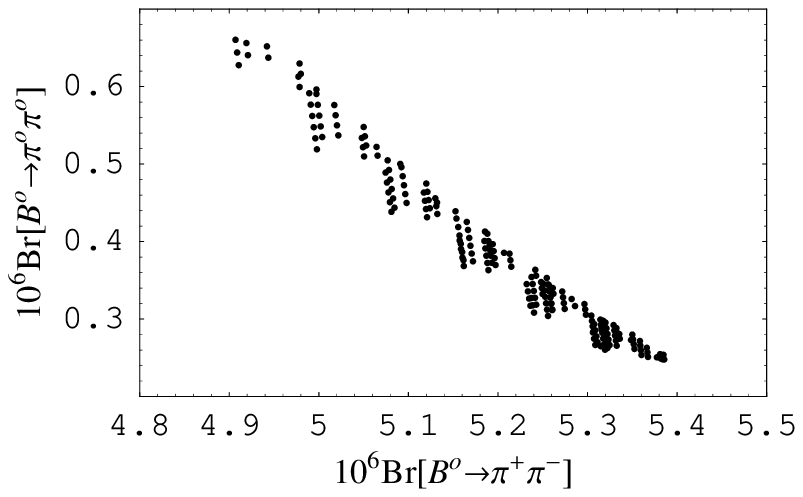}%
}%
\phantom{asdfghj} {\includegraphics[ height=1.3612in,
width=2.1906in
]%
{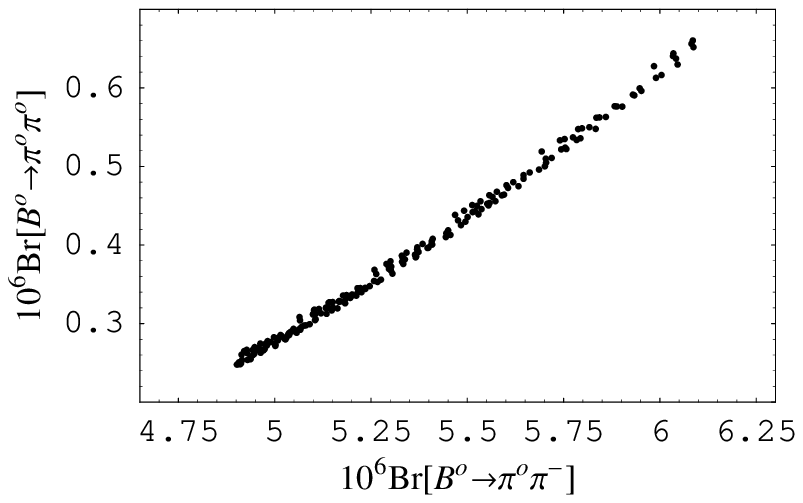}%
}
\caption{ Branching fractions  and and  amplitudes
$\hat\alpha_{1,2}=\alpha_{1,2}/f_0$  as functions of hadron input
parameters \ $f_{B}(0.02),~\lambda_{B}(0.06),~a_{2}^{\pi}$(0.06)
and $a_{4}^{\pi}$(0.06). The numbers in the brackets give the
value of step in the numerical calculations.
 The large points in the upper plots
correspond to the choice
$f_{B}=0.23,~\lambda_{B}=0.23,~a_{2}^{\pi}=0.30$ and
$a_{4}^{\pi}=-0.07$. The values of the $\hat \alpha_i$ and
branching fractions which lie outside of
experimental interval  are not shown.}%
\label{Nres}%
\end{center}
\end{figure}

For the B-meson LCDA we accept a simple model with
exponential behavior which is very popular in phenomenological applications%
\[
\phi_{B}(\omega)=\frac{\omega}{\lambda_{B}^{2}}\exp\left(  -\omega/\lambda
_{B}\right)  .
\]
Then one can easily  calculate  the moments introduced in
eq.(\ref{phiBmom}):%
\[
\left\langle L\right\rangle =\ln\left[  \frac{m_{b}\lambda_{B}}{\mu_{hc}^{2}%
}\right]  -\gamma_{E},~~\left\langle L^{2}\right\rangle =\ln^{2}\left[
\frac{m_{b}\lambda_{B}}{\mu_{hc}^{2}}\right]  -\gamma_{E}+\frac{\pi^{2}}{6},
\]
where $\gamma_{E}=0.577...~$. With the given above central  value of $\lambda_{B}$ one
obtains%
\[
\left\langle L\right\rangle =-0.87,~\ \ ~\left\langle L^{2}\right\rangle =2.4
\]
For QCD running coupling $\alpha_{S}$ we use the two loop approximation with
QCD scale $\Lambda_{QCD}^{(5)}=225$MeV. Recall, that in our numerical estimates
of the branching fractions we neglect  EW-penguins contributions
but include QCD
penguins in the NLO approximation as given in \cite{BBNS}. As it was
observed in those papers, the values of the pion branching fractions are very
sensitive to the product \ $\left\vert V_{ub}\right\vert f_{0}$ . Corresponding
value can be estimated from semileptonic decay $B\rightarrow\pi l\nu$
assuming monotonic behavior of
the form factor $f_{+}(q^{2}):$%
\[
\frac{d\Gamma(B^{0}\rightarrow\pi l\nu)}{dq^{2}}=\frac{G_{F}^{2}}{24\pi^{2}%
}\left\vert V_{ub}\right\vert ^{2}\left\vert f_{+}(q^{2})\right\vert
^{2}~p_{\pi}^{3}=\frac{G_{F}^{2}}{24\pi^{2}}\left\vert V_{ub}\right\vert
^{2}\left\vert f_{+}(q^{2})\right\vert ^{2}~p_{\pi}^{3}>\frac{G_{F}^{2}}%
{24\pi^{2}}\left\vert V_{ub}\right\vert ^{2}\left\vert f_{0}(0)\right\vert
^{2}~p_{\pi}^{3}.
\]
Using results obtained by BABAR collaboration in \cite{BABARf0} for the lowest bin in
$q^{2}<8$GeV$^{2}:$%
\[
\tau_{B}\int_{0}^{8}dq^{2}\frac{d\Gamma(B^{0}\rightarrow\pi l\nu)}{dq^{2}%
}=\Delta Br(B^{0}\rightarrow\pi l\nu)=0.21\pm0.13
\]
$~$one obtains%
\[
10^{3}\left\vert V_{ub}\right\vert f_{0}<\sqrt{\frac{\Delta Br(B^{0}%
\rightarrow\pi l\nu)}{\frac{\tau_{B}G_{F}^{2}}{24\pi^{2}}\int_{0}^{8}%
dq^{2}p_{\pi}^{3}}}\times10^{3}=0.72_{-0.27}^{+0.20}~(1.0),
\]
where in brackets we show the product of the central values $\left\vert
V_{ub}\right\vert $ and $\ f_{0}$ from the Table 1. \ In order to satisfy
this requirement (at least for upper bound) we accept following values for
$\left\vert V_{ub}\right\vert $ and $\ f_{0}:$%
\begin{equation}
\left\vert V_{ub}\right\vert =0.0038,~~f_{0}=0.23,~~\text{with}%
~~10^{3}\left\vert V_{ub}\right\vert f_{0}=0.87~~. \label{Vubf0}%
\end{equation}
As one can
see from Table~1, such choice of the $\left\vert V_{ub}\right\vert $ and $\ f_{0}$  corresponds to the lowest possible value of $\ f_{0}$ within indicated
\ uncertainties. First,  we compute two largest
branching fractions as a
functions of four parameters $f_{B},~\lambda_{B},~a_{2}^{\pi}$ and $a_{4}%
^{\pi}$. These results for the tree level dominant branchings
 $Br(B\rightarrow\pi^{-}%
\pi^{+})$ and $Br(B\rightarrow\pi^{-}\pi^{0})~\ $\ and corresponding values
for $\alpha_{1,2}$ are presented in Fig.\ref{Nres}. We show all solutions
which   describe the
experimental points changing the parameters inside the
intervals indicated in the Table~1. As one can observe, there exist
many possible
solutions that demonstrate large ambiguity due to badly
known mesons parameters.
\ For instance, we reproduce the experimental values
\begin{align}
10^{6}Br(B  &  \rightarrow\pi^{-}\pi^{+})=5.1~(\text{exp: }5.0\pm0.4),\\
10^{6}Br(B  &  \rightarrow\pi^{-}\pi^{0})=5.51~(\text{exp: }5.5\pm0.6),
\end{align}
with $f_{B}=0.23,~\lambda_{B}=0.23,~a_{2}^{\pi}=0.3$ and $a_{4}^{\pi}=-0.07$.
\ Corresponding amplitudes $\alpha_{1,2}$ have following numerical  structure
at this point:%
\begin{eqnarray}
\alpha_{1}/f_{0}  &  = &\left[  1.04+0.012i\right]  _{V}+  \left(
-0.030\right)  _{T^{LO}\ast J^{LO}}+\left(  -0.020\right)  _{T^{LO}\ast
J^{NLO}}+\left(  -0.035-0.031i\right)_{T^{NLO}\ast J^{LO}}
\nonumber \\
&=&0.96-0.019i,
\label{alfa1num}\\[0.2cm]
\alpha_{2}/f_{0}  &  =&\left[  0.035-0.077i\right]  _{V}+
 \left(
0.19\right)  _{T^{LO}\ast J^{LO}}+\left(  ~0.13\right)  _{T^{LO}\ast J^{NLO}%
}+\left(  ~0.028+0.060i\right)  _{T^{NLO}\ast J^{LO}}  %
\nonumber \\ \label{alfa2num}%
&&
=0.38-0.020i,
\end{eqnarray}
where  for convenience we presented the answers normalized to $1/f_{0}$. Let us
briefly comment these results. \ The real part of the $\alpha_{1}$ is clearly
dominated by the vertex term, the corrections from the hard spectator scattering
are about $5\%$ in absolute size. As one can see from (\ref{alfa1num}) the
radiative corrections ( indicated as $_{T^{LO}\ast J^{NLO}}$ and
$_{T^{NLO}\ast J^{LO}}$) numerically quite large with respect to
LO term $_{T^{LO}\ast J^{LO}}$. The
relatively large value of  $T^{NLO}\ast J^{LO}$ contribution is explained by
large value of the Wilson coefficient $C_{1}$ with respect to $C_{2}$.

For the amplitude
$\alpha_{2}$ the situation for the real part is different. The vertex
contribution is very small due to the compensation between tree and NLO
contributions \cite{BBNS}.
Therefore the dominant term arises from the hard spectator scattering part of
the amplitude. The structure of the NLO terms here is also different: bulk of
the radiative correction  is due to NLO jet function. The NLO hard
spectator scattering is approximately  four times smaller.
Hence obvious conclusion is
that the total (jet+hard) NLO contribution is very important for
 Re[$\alpha_{2}$] and almost
negligible for Re[$\alpha_{1}$]  respectively.

The important result of the our calculation is the  imaginary part of
 both amplitudes $\alpha_{1,2}$.  Its value,  in comparison with the value of
imaginary part from the vertex contribution $V$, is quite large
and has opposite sign. Therefore the resulting  imaginary parts are
significantly modified. Such changes may produce sizable effect for the
CP-violating asymmetries and therefore have to be taken into account in
 phenomenological analysis.

The smallness of the $\alpha_{2}$ provides small value of the
third branching $10^{6}Br(B\rightarrow\pi^{0}\pi^{0})$. Its value always
remains  considerably smaller than the experimental value, see
two bottom plots in Fig~(\ref{Nres}). With the amplitude
$\alpha_2$ from
eq.(\ref{alfa2num}) we obtained%
\[
10^{6}Br(B\rightarrow\pi^{0}\pi^{0})=0.45~(\text{exp: }1.45\pm0.29)
\]
that is three times smaller then the experimental value. Of course, there exist
ambiguities not only due to hadronic input parameters but also in the
 scale setting,
 in quark masses and weak parameters. Such uncertainties
have been already estimated in \cite{Beneke05} and they are quite large.

On the other side it's possible to suppose that  realistic explanation of the
small theoretical value of the $\pi^{0}\pi^{0}$ branching can be explained
by relatively large
contributions of the power corrections which have been ignored in present
calculations.
The key observation is that $B\rightarrow\pi^{0}\pi^{0}$ amplitude has small
absolute value ($\sim \alpha_{S}$)  due to  cancellation among the
vertex contributions. Then it makes possible that \ preasymptotic
effects from the power corrections are very important especially for this
case.
In  papers \cite{BBNS,Beneke05}
 the model for some contributions of the power corrections have been
already introduced to estimate their effect.
In particular, the so-called  "chiral enhanced" twist-3 contributions
have been considered as
a source of dominant effect. In \cite{Beneke05} such correction  strongly
enhances the absolute value of $\alpha_2$ compare to  perturbative contribution.

Similar situation, with small leading power term and large power suppressed
corrections occurs in hard exclusive processes.
 Such scenario, as expected, is realized for the pion form factor
in large $Q^{2}$ limit. In that situation leading twist perturbative
contribution is small $~\sim O(\alpha_{S})$  and power correction may even
dominate in quite large accessible range of $~Q^{2}$. For the detailed discussion
of this question we refer to \cite{Braun1995} where light-cone
sum rule approach have been used for analysis of the power behavior in $Q^2$.
 The other interesting observation, which was made in \cite{Braun1995},
is related to the "chiral enhanced" contributions. For the pion
form factor such corrections naively could provide very strong
effect as it was  observed first in \cite{Terentev}. But in  sum
rule calculations it was found that such contributions turn out
to be small. It might be understood, that the value  of such
corrections is overestimated if one uses the simple model with a
cutoff of the momentum fraction to ensure convergence of the
convolution integrals. If this true, then for description of
$B\rightarrow\pi\pi$ decays one needs, probably, a different model of the
power corrections than one  used up to now. The detailed discussion
of this question lies beyond the scope of present consideration and we refer
to recent works dedicated to this subject
\cite{Feldmann:2004mg}.

\bigskip

\section{Conclusions}

We have presented the independent calculation of the next-to-leading
order corrections to the graphical tree amplitudes in $B\rightarrow\pi\pi$
decays. Our analytical expressions are in agreement with
results obtained in \cite{Beneke05} using a  different technical approach.
The obtained results have been used for the numerical estimates of the
branching fractions in BBNS approach \cite{BBNS}.
We have found that total ( hard plus jet ) next-to-leading
correction is relatively
small for the real part of the $\alpha_1$ decay amplitude and provide
large contribution to the real part of $\alpha_2$. In the last case the
dominant effect originate from the next-to-leading contribution of the
jet function. But the imaginary  part which is generated by the hard coefficient
function of the hard spectator scattering term is quite large and
therefore can provide sizable contribution to  the CP-violating asymmetries.
Our estimates of the branching fractions allows to make conclusion about
existence of sizable effect from power suppressed contributions, especially for
branching $B\rightarrow\pi^0\pi^0$ which dominated small $\alpha_2$ amplitude.
\bigskip

\subsection*{Acknowledgments}

We are grateful to M.\ Beneke and S. J\"{a}ger for very
interesting discussions.
 This work is supported by Sofia Kovalevskaya Program of the Alexander von
 Humboldt Foundation.

\bigskip

\section*{Appendix }
\subsection*{1. Structure of the divergent contributions of
the Feynmann integrals}

Here  we briefly discuss the structure of different divergent contributions and
provide results for the singular parts of the $UV-$ and $IR-$ integrals.

The $UV-$divergencies arising in the NLO diagrams are removed by the
 counterterms for the four-quark operators, renormalization constants of the
 wave functions and QCD counterterms. Renormalization of the four fermion operators
 $O_{1,2}$ is given by:
\begin{align}
O_{1}  &  =Z_{11}Z_{\psi}^{-2}O_{1}^{bare}+Z_{12}Z_{\psi}^{-2}O_{2}%
^{bare},~\ \\
O_{2}  &  =Z_{12}Z_{\psi}^{-2}O_{1}^{bare}+Z_{22}Z_{\psi}^{-2}O_{2}^{bare}%
\end{align}
where~$2\times2$ leading order matrix $Z~$\ reads (see for instance \cite{Buras}):
\begin{equation}
Z=1-\frac{1}{\varepsilon}\frac{\alpha_{s}}{4\pi}\left(
\begin{array}
[c]{cc}%
2C_{F}-6(C_{F}-C_{A}/2) & -3\\
-3 & 2C_{F}-6(C_{F}-C_{A}/2)
\end{array}
\right)  \label{Z}%
\end{equation}
and $Z_{\psi}$ is the renormalization constant of the quark field in the
$\overline{\text{MS}}$-scheme.%
\begin{equation}
Z_{\psi}=1-\frac1\epsilon\frac{\alpha_{s}}{4\pi}C_{F}.
\end{equation}
In addition,  matrix elements have to be multiplied on the \ corresponding
renormalization constant of the wave functions external particles.
For the quark wave function
such renormalization factor is defined by%
\begin{equation}
Z_{q}=1+\left.  i\frac{d\Sigma_{\psi}}{d\hat{p}}\right\vert _{\hat{p}=p},
\label{def:Zq}
\end{equation}
where we assume off-shell $IR-$regularization and $\Sigma_\psi$ denotes one-particle
irreducible self-energy graphs. Such definition introduces, except $UV-$pole,
 the finite term:
\begin{equation}
\frac{1}{2}(Z_{q}-1)=-\frac1\epsilon\frac{\alpha_{s}}{4\pi}C_{F}
-\frac{\alpha_{s}}{4\pi}C_{F}\left[
\ln\frac{\mu^{2}_R}{p^2}+Z_{fin}
\right],
\label{ZqR}
\end{equation}
The pole part cancel  $UV-$divergencies arising in diagrams for the matrix element.
The finite term must be included in the matching.
But the  same  definition (\ref{def:Zq}) is used
for the renormalization of the quark wave functions in the effective theory.
One obtains similar contribution:
\begin{equation}
\frac{1}{2}(Z_{q}-1)_{\rm fin}=
-\frac{\alpha_{s}}{4\pi}C_{F}\left[
\ln\frac{\mu^{2}_F}{p^2}+Z_{fin}
\right],
\label{ZqF}
\end{equation}
which is different from (\ref{ZqR}) only by renormalization scale $\mu^{2}_F$. Hence,
the difference of  two expressions (\ref{ZqR}) and (\ref{ZqF}) which defines
the contribution to the hard coefficient function
 from such terms is proportional to $\ln(\mu_{R}%
/\mu_{F})$. We put $\mu_{R}=\mu_{F}$ that allows to avoid consideration of such
terms. The same arguments can be repeated for the gluon wave functions.
But situation with heavy $b$-quark is different. The HQET wave
function of the effective filed $h_{v}$ is renormalized by factor $Z_{h}$%
\begin{equation}
Z_{h}=1+\left.  i\frac{d\Sigma_{h}}{d(vk)}\right\vert _{(vk)=0},
\end{equation}
where $k$ denotes residual momentum. Then one obtains:%
\begin{align}
\frac{1}{2}\left(  Z_{b}-1\right)   &  =\frac{\alpha_{S}}{4\pi}C_{F}\left(
-\frac{1}{\varepsilon}-2\ln\left(  \frac{2(vk)}{\mu_{R}}\right)  -3\ln\left(
\frac{\mu_{R}}{m_{b}}\right)  -2\right)  ,\\
\frac{1}{2}\left(  Z_{h}-1\right)   &  =\frac{\alpha_{S}}{4\pi}C_{F}\left(
-\frac{1}{\varepsilon}-2\ln\left(  \frac{2(vk)}{\mu_{F}}\right)  \right)  ,
\end{align}
Hence the corresponding contribution to the hard coefficient function \ (
for $\mu_{R}=\mu_{F}$)
\begin{equation}
\left(  T_{i}\right)  _{\Sigma_{h}}=\frac{\alpha_{S}}{2\pi}C_{F}\left(
\frac{C_{i\pm1}}{N_{c}}\frac{1}{\bar{u}}\right)  \left(  -1-\frac{3}{2}%
\ln\frac{\mu}{m_{b}}\right)  .
\end{equation}

After these remarks let us provide list of the singular contributions for the
integrals $(J_{X})_{UV}$ defined in \ (\ref{def:UVJ1}) and  for the soft and
collinear integrals (\ref{Jregions}),(\ref{def:collsoft})
which appear in graphs presented in Fig.\ref{Fig:NLOgraphs}.
For simplicity, we shall indicate below
index $IR$ for the soft and collinear integrals and define ($X=A1,A2,...$)%
\begin{equation}
{\rm pole\, \, terms }\left[\left(J_{X}\right)_{hard}\right]=
-{\rm pole\,\, terms }\left[\left(J_{X}\right)_{soft,coll}\right]=X_{col} JX_{IR}
\label{def:JXIR}
\end{equation}
where $X_{col}$  denotes color factor as in (\ref{def:UVJ1}).
The explicit expressions for $X_{col}$  are listed in Table1.

\begin{center}
Table1. The color factors of different graphs in~Fig.~\ref{Fig:NLOgraphs}.\\%
\begin{tabular}
[c]{|l|l|}\hline
 & \\[-0.3cm]
Color factors $X_{col}$ & diagram index  $X$
\\
& \\[-0.3cm]\hline
 & \\[-0.3cm]
$\frac{C_{i}}{2N_{c}}+\frac{C_{i\pm1}}{N_{c}}\left(  C_{F}-\frac{C_{A}}%
{2}\right)  $ & $A1,~B6,~C1,~D6$
\\
& \\[-0.3cm]\hline
 & \\[-0.3cm]
$\frac{C_{i}}{2N_{c}}+\frac{C_{i\pm1}}{N_{c}}C_{F}$ & $A6,~C6,~D1$
\\
& \\[-0.3cm]\hline
 & \\[-0.3cm]
$\frac{C_{i\pm1}}{N_{c}}\left(  C_{F}-\frac{C_{A}}{2}\right)  $ &
$A2,C2,D2,E6,F6,H1$
\\
& \\[-0.3cm]\hline
 & \\[-0.3cm]
$\frac{C_{i\pm1}}{N_{c}}C_{F}$ & $A4,A7,B4,C7,D4,C4,E1,E2,H6$
\\
& \\[-0.3cm]\hline
 & \\[-0.3cm]
$\frac{C_{i\pm1}}{N_{c}}\frac{C_{A}}{2}$ & $A3,C3,D3,E3,H3$%
\\[-0.3cm]
& \\ \hline
\end{tabular}
\end{center}
\noindent
As usually we use notation $\bar {u}\equiv1-u$ \ and $\overline{\text{MS}%
}$--scheme for subtractions. For simplicity we put $\mu=m_{b}$. Then%
\begin{align}
JA6_{UV}  &  =JC6_{UV}=JE6_{UV}=JF6_{UV}=JH1_{UV}=-JH6_{UV}=-\frac
{1}{2\bar {u}\varepsilon},~JH3_{UV}=-\frac{3}{2\bar {u}\varepsilon
},~~\ \\
JA4_{UV}  &  =JC4_{UV}=-\frac{1}{2\varepsilon},
 ~JB4_{UV}=JD4_{UV}=\frac{2}{\varepsilon}%
,~JB6_{UV}=JD6_{UV}=\frac{2}{\varepsilon\bar {u}},
\end{align}%
\begin{equation}
JA1_{IR}=\frac{1}{\bar {u}}\left(  \frac{1}{\varepsilon^{2}}+\frac
{1}{\varepsilon}\left(  1+i\pi-\ln[z\bar {u}]\right)  \right)
,~~JA2_{IR}=\frac{1}{\varepsilon}\frac{1}{\bar {u}}\left(  -\frac{1}%
{2}-\frac{z\ln z}{\bar {z}}\right)  ,
\end{equation}%
\begin{equation}
JA3_{IR}=\frac{1}{\varepsilon}\frac{1}{\bar {u}}\left(  1+\ln\bar
{z}\right)  ,~~JA4_{IR}=\frac{1}{\varepsilon^{2}}+\frac{1}{\varepsilon}\left(
2+i\pi-\ln[z\bar {u}]\right)  ,~JA7_{IR}=-\frac{1}{\varepsilon}\frac
{1}{2\bar {u}},
\end{equation}%
\begin{equation}
JB4_{IR}=-\frac{1}{\varepsilon^{2}}+\frac{1}{\varepsilon}\left(
-2-i\pi +\ln[zu]\right)  ,~~JB6_{IR}=\frac{1}{\bar {u}}\left(
-\frac {1}{\varepsilon^{2}}-\frac{1}{\varepsilon}\left(
2+i\pi-\ln[zu]\right) \right)  ,
\end{equation}%
\begin{equation}
JC2_{IR}=\frac{1}{2\varepsilon^{2}}-\frac{1}{\varepsilon}\left(  \ln
u-\frac{1-z\bar {u}}{uz}\ln\left[  \frac{1-z\bar {u}}{1-z}\right]
\right)  ,
\end{equation}%
\begin{equation}
JC3_{IR}=\frac{1}{\varepsilon^{2}}+\frac{1}{\varepsilon}\left(  -\frac{1}%
{2}+i\pi-\ln u-\frac{\bar {z}+2uz}{uz}\ln\bar {z}+\frac{1-z\bar
{u}}{zu}\ln\left[  1-z\bar {u}\right]  \right)  ,
\end{equation}%
\begin{equation}
JC4_{IR}=\frac{1}{\varepsilon}\left(  1-\frac{1-z\bar {u}}{zu}\ln\left[
\frac{1-z\bar {u}}{\bar {z}}\right]  \right)  ,~JC6_{IR}=\frac
{1}{\bar {u}}\left(  \frac{1}{2\varepsilon^{2}}+\frac{1}{\varepsilon
}\left(  1-\ln u\right)  \right)  ,
\end{equation}%
\begin{equation}
JD1_{IR}=\frac{1}{\bar {u}}\left(  -\frac{1}{2\varepsilon^{2}}+\frac
{1}{\varepsilon}\ln\bar {u}\right)  ,~~JD2_{IR}=-\frac{1}{2\varepsilon
^{2}}+\frac{1}{\varepsilon}\left(  \ln\bar {u}-\frac{1-uz}{z\bar {u}%
}\ln\left[  \frac{1-uz}{\bar {z}}\right]  \right)  ,
\end{equation}%
\begin{equation}
JD3_{IR}=\frac{1}{\varepsilon^{2}}\left(  \frac{1}{2\bar
{u}}-1\right) -\frac{1}{\varepsilon}
\left(  i\pi-\frac{u}{2\bar{u}}-\ln\bar {u}-
\frac{(1-2uz)}{\bar {u}z}\ln\bar{z}+
\frac{(1-uz)}{\bar {u}z}\ln[1-uz]
\right)  ,
\end{equation}%
\begin{equation}
JD4_{IR}=\frac{1}{\varepsilon}\left(  -1+\frac{1-uz}{\bar {u}z}\ln
\frac{1-uz}{\bar {z}}\right)  ,~~JE2=-\frac{1}{\varepsilon}\frac
{\ln\bar {u}}{u},
\end{equation}%
\begin{equation}
JE3_{IR}=\frac{1}{\bar {u}}\left(  -\frac{1}{\varepsilon^{2}}+\frac
{1}{\varepsilon}\left(  -\frac{1}{2}-i\pi+\frac{\ln\bar {u}}{u}+\ln\left[
u~\bar {z}\right]  \right)  \right)  ,
\end{equation}%
\begin{equation}
~JE6_{IR}=\frac{1}{\varepsilon}\frac{1}{\bar {u}}\left(  1+\frac
{\ln\bar {u}}{u}\right)  ,~JF6_{IR}=\frac{1}{\bar {u}}\left(  \frac
{1}{2\varepsilon^{2}}+\frac{1}{\varepsilon}\left(  1-\ln z\right)  \right)  ,
\end{equation}%
\begin{equation}
JH1_{IR}=\frac{1}{\varepsilon}\frac{1}{\bar {u}},~~JH3_{IR}=\frac
{1}{\bar {u}}\left(  \frac{1}{\varepsilon^{2}}+\frac{1}{\varepsilon
}\left(  \frac{3}{2}+i\pi-\ln\left[  \bar {z}\bar {u}\right]  \right)
\right)  .
\end{equation}
Performing analysis of the main regions which  contribute to the leading
power accuracy we find that many $IR-$contributions cancel in certain
combinations of diagrams. Important
that  such cancellation  can be observed  before computing of the integrals
and therefore provides
a good check for the intermediate calculations. Taking
into account the color factors of diagrams we have found following $IR-$finite
combinations:%
\begin{equation}
\left(  JA1+JB6+JC6+JD1\right)  _{IR}=0,~~\left(
JA4+JB4+JC2+JC4+JD2+JD4\right)  _{IR}=0,
\end{equation}%
\begin{equation}
\left(  JA1+JB6-JE3+JE6+JH1-JH3\right)  _{IR}=0.
\end{equation}
The remaining $IR-$contributions can be associated with the matrix elements of the
SCET operators which define LCDA $\varphi_{P}^{NLO}$ and  from factor $\theta^{NLO}$.
To obtain results for the $Z_{0,1,2}$ introduced in (\ref{Z}) we have
computed following expressions:%
\begin{align}
\left[  \frac{1}{\varepsilon}Z_{1}+Z_{0}\right]  _{coll-p^{\prime}}  &
=\frac{C_{i\pm1}}{N_{c}}C_{F}\left(  \left(  JE2+JE6+JH1\right)
_{coll-p^{\prime}}+\int_{0}^{1}dx^{\prime}\frac{\varphi_{P}^{NLO}(x^{\prime}%
)}{1-x^{\prime}}\right) \nonumber \\
&  =\frac{C_{i\pm1}}{N_{c}}C_{F}\left(  \frac{1}{\varepsilon}\frac
{1}{\bar {u}}(2+\ln\bar {u})+\frac{\ln\bar {u}}{u}\right)  ,
\label{phi:pole}
\end{align}%

\begin{align}
\left[  \frac{1}{\varepsilon^{2}}Z_{2}+\frac{1}{\varepsilon}Z_{1}%
+Z_{0}\right]  _{s/coll-p}=\frac{C_{i\pm1}}{N_{c}}\left[  C_{F}\left(
JA2+JA7+JF6\right)  _{s/coll-p}\right.  + \nonumber
\\
\left.  \frac{C_{A}}{2}\left(  JA3-JA2-JC2+JC3-JD2+JD3-JF6\right)
_{s/coll-p}+\int_{0}^{1}dz^{\prime}\theta^{NLO}(z^{\prime})\right]  =
\nonumber
\\
\frac{C_{i\pm1}}{N_{c}}\left[  C_{F}\frac{1}{\bar {u}}\left(  \frac
{1}{2\varepsilon^{2}}-\frac{1}{\varepsilon}\frac{\ln z}{\bar {z}}-\frac
{1}{2}-\frac{\ln z}{\bar {z}}\right)  +\frac{C_{A}}{2}\frac{1}%
{\bar {u}}\left(  \frac{1}{\varepsilon}\frac{\bar {z}+\ln z}%
{\bar {z}}+1+\frac{\ln z}{\bar {z}}+\frac{\ln\bar {z}}{z}\right)
\right]
\label{thet:pole}
\end{align}

The pole contribution in formula (\ref{phi:pole}) can be interpreted as
convolution of the evolution kernel
with the leading order coefficient function
$T_i^{LO}$:
\begin{equation}
(Z_{1})_{coll-p^{\prime}}+\tilde Z_{\psi}=
V  \ast\widetilde{t}_{i}^{LO}\ast\theta^{LO}=C_{F}\left(
-\frac{C_{_{i\pm1}}}{N_{c}}\right)  \frac{1}{\bar {u}}\left(
\frac32+\ln\bar {u}\right) ,
\label{Z1p}
\end{equation}
where in the left side of eq.(\ref{Z1p}) we introduced
contribution from the quark field renormalization,
denoted as $\tilde Z_{\psi}$. We added this term because the set  diagrams
in (\ref{phi:pole}) doesn't
 have such contribution and therefore corresponding poles define  non-trivial
but not complete part of the evolution kernel $V$:
\begin{align}
V(x,u)  &  =C_{F}\left[  \frac{x}{u}\theta(x<u)\left(  1+\frac{1}{u-x}\right)
+\frac{1-x}{1-u}\theta(x>u)\left(  1+\frac{1}{x-u}\right)  \right]  _{+}%
\end{align}
where  plus-prescription denotes:
$\left[  F(x,u)\right]  _{+}=F(x,u)-\delta(x-u)\int_{0}^{1}dx^{\prime
}F(x^{\prime},u).$
The same consideration can be carried out for the poles in (\ref{thet:pole}).
Because leading order coefficient functions $T_{i}^{LO}$ \ (\ref{TLOi}) are
independent on the momentum fraction $z^{\prime}$, the singular (pole) part
 of the
form factor $\theta^{NLO}(z')$ appears as integral $\int dz^{\prime
}~\theta^{NLO}(z^{\prime})$ and can be understood  as counterterm of the
local operator $J^{B1}(s=0)$ in the effective theory. We have checked by
independent calculation that expressions for the $Z_{1,2}$ in (\ref{thet:pole}) is
in agreement with the renormalization  of the local  \scetI operator $J^{B1}(s=0)$
(\ref{JB1}):
\begin{align}
\frac{1}{\varepsilon^{2}}Z_{2}+
\frac{1}{\varepsilon}\left[(Z_{1})_{s/coll-p}+\tilde Z_{\psi}+
 \frac12\tilde Z_A+\tilde Z_g \right] &
=\varphi_{P}^{LO}\ast T_{i}^{LO}\ast(\theta^{NLO})_{UV-pole}%
=\nonumber\\
&  \left(  -\frac{C_{_{i\pm1}}}{N_{c}}\frac{1}{\bar {u}}\right)  \left(
-\frac{1}{\varepsilon^{2}}\frac{C_{F}}{2}(1+2\varepsilon\ln\left[  \mu
/(p\bar n)\right]  )\right. \nonumber\\
&  ~~~~~\ ~\ \ \ \ ~~\ \left.
-\frac{1}{\varepsilon}\left[  C_{F}\frac{\bar z-4\ln
z}{4(1-z)}+\frac{C_{A}}{2}\frac{\ln z}{1-z}\right]  \right)\, ,%
\label{Z21soft}
\end{align}
where we again introduced contribution from the renormalization
factors for coupling $\tilde Z_g $ and fields $\tilde Z_{\psi ,A}$
which necessary for complete definition of the evolution kernel.
Equation (\ref{Z21soft}) is in agreement with known results for
the evolution $J^{B1}$ obtained in \cite{BY05,NBH}.

\subsection*{ 2. Finite contributions of the hard integrals
$ (J_X)_{hard}$} %

In the second part of Appendix we present the finite contributions of the
hard integrals $ (J_X)_{hard}$ introduced
 in (\ref{Jregions}) and (\ref{def:Jhard}). We shall write
 \begin{equation}
{\rm finite\, \, terms }\left[\left(J_{X}\right)_{hard}\right]=X_{col} JX
\label{def:JX}
\end{equation}
 As usually, we assume $\mu=m_b$ in order to simplify the formulas.
\begin{align}
JA1  &  =\frac{2}{\bar {u}}-\frac{7{\pi}^{2}}{12\,\bar {u}}-\frac
{\ln\bar {u}}{\bar {u}}+\frac{\left(  1+z\right)  \,\ln\bar {z}%
}{2\,\bar {u}\,z}+\frac{\left(  3\,z-2\right)  \,\ln z}{2\bar
{u}\,\bar {z}}
 +\frac{{\ln}^{2}({z~}\bar {{u}}{)}}{2\,\bar {u}}+\frac{{i\pi
}\,\left(  1-\ln[\bar{u}\,z]\right)  }{\bar {u}}\,~,
\end{align}

\begin{equation}
JA2=-\frac{1}{\bar {u}}+\frac{\ln\bar {u}}{2\bar {u}}+\left(
-\frac{3\,z}{2}+z\,\ln\bar {u}\right)  \frac{\,\ln z}{\bar
{u}\,\bar {z}}+\frac{z\,{\ln}^{2}{z}}{2\,\bar {u}\,\bar {z}}~
-i\pi \left(
\frac{1}{2\bar u}+\frac{z\ln z}{\bar u\bar z}
\right),
\end{equation}

\begin{align}
JA3  &  =\frac{2}{1-u}-\frac{\ln(1-u)}{1-u}+\frac{\ln(1-z)}{2\,\left(
1-u\right)  \,z}-\frac{\ln(1-u)\,\ln(1-z)}{1-u}-
\frac{{\ln}^{2}{(1-z)}}{2\,\left(  1-u\right)  }+\frac{{i\pi}\,\left(
1+\ln\bar {z}\right)  }{1-u}~,
\end{align}

\begin{equation}
JA4=4-\frac{7{\pi}^{2}}{12}-\frac{3\ln[\bar {u} z]}{2}%
+\frac{{\ln}^{2}[{z~}\bar {{u}}{)]}}{2}+{i\pi}\,\left(  \frac{3}{2}%
-\ln(\bar{u}\,z)\right)  ~,
\end{equation}

\begin{equation}
JA6=-\frac{{i\pi}}{2\,\bar {u}}-\frac{\ln\bar {z}+z-z\ln(\bar{u}%
\,\bar{z})}{2\,\bar {u}\,z}~,
\end{equation}

\begin{equation}
JA7=-\frac{{i\pi}}{2\,\bar {u}}-\frac{2-\ln\bar{u}}{2\,\bar {u}}~,
\end{equation}

\begin{equation}
JB4=  {i\pi}\,\ln[u\,z]+\frac{7{\pi}^{2}}{12}+
\frac{1}{2}(3-\ln^2[u z]) ~,
\end{equation}

\begin{equation}
JB6=\frac{1}{\bar {u}}\left(  {i\pi}\,\ln(u\,z)+\frac{1-{\log}^{2}%
{[u\,z]}}{2}+\frac{7{\pi}^{2}}{12}\right)  ~,
\end{equation}

\begin{equation}
JE1=-\frac{\ln u}{2\,\bar {u}}~,
\end{equation}

\begin{equation}
JE2=\frac{1}{2u}\left(  {-2\,{i\pi}\,\ln\bar {u}-3\ln\bar {u}+{\ln}%
}^{2}\bar {u}{+2\,\ln\bar{u}\,\ln\bar{z}}\right)  ~,
\end{equation}

\begin{align}
JE3  &  = -\frac{1}{{\bar {u}}}+\frac{7\pi^{2}}{12{\bar {u}}%
}-\frac{{\ln}^{2}{\bar {u}}}{2u{\bar {u}}}-\frac{\ln^{2}%
[u\bar z]}{2{\bar {u}}}+\frac{\ln{\bar {z}}}{2{\bar {u}}}+
\frac{3}{2{\bar {u}}}\ln u~
+\ln\bar {u}\left(  \frac{3-2u}{2u{\bar {u}}%
}-\frac{\ln{\bar {z}}}{u{\bar {u}}}\right)  -\nonumber\\
&  \left.  {i\pi}\,\frac{\left(  u-2\,\ln{\bar {u}}-2\,u\,\ln[u\,\bar
{z}]\right)  }{2u{\bar {u}}}~\right)  ~,
\end{align}

\begin{equation}
JE6=\frac{1}{2\,{\bar {u}}\,u}\left(  4\,u+\left(  3-u\right)
\,\ln{\bar {u}}-{\ln}^{2}{\bar {u}}-u\,\ln\bar {z}-2\,\ln
{\bar {u}}\,\ln{\bar {z}}+{i\pi}\,\left(  u+2\,\ln\bar{u}\right)
\right)  ~~,
\end{equation}

\begin{equation}
JF6=\frac{\pi^{2}}{24{\bar {u}}}+\frac{2}{{\bar {u}}}-\frac{\left(
2-3\,z\right)  \,\ln z}{2\,{\bar {u}}\bar {{z}}}+\frac{1}%
{{\bar {u}}}\text{S}\left(  \frac{\bar {z}}{z}\right)  ~,
\end{equation}

\begin{equation}
JH1=\frac{1}{2\,{\bar {u}}}\left(  4
+{i\pi}-\ln(\bar{u}\,\bar{z})\right) ~,
\end{equation}

\begin{equation}
JH3=-\frac{7{\pi}^{2}}{12\,{\bar {u}}}+\frac{{\ln}^{2}{(\bar{u}\,\bar{z})}%
}{\,{2\bar {u}}}-\,{i\pi}\frac{\,\ln(\bar{u}\,\bar{z})}{\,{\bar {u}}}~,
\end{equation}

\begin{equation}
JH6=\frac{1}{2\,\,{\bar {u}}}\left(  \ 1+{i\pi}-\ln(\bar{u}\,\bar
{z})\ \right)  ~,
\end{equation}

\begin{align}
JC1  &  =\mathrm{Re}JC1+i\pi\mathrm{Im}JC1,\\
\mathrm{Im}JC1  &  =\frac{\bar {z}\,\left(  1-3\,u-z\right)  }{2\,{\left(
1-u-z\right)  }^{2}}-\frac{u^{2}\,\bar {z}}{{\left(  1-u-z\right)  }^{3}%
}\,\ln\left[  \frac{u}{\bar {z}}\right]  ,
\end{align}%
\begin{align}
\mathrm{Re}JC1  &  =-\frac{u^{2}\,\bar{z}\,\,\mathrm{I}(u,\bar{z})}{{\left(
1-u-z\right)  }^{2}}-\frac{\ln u}{2}\left(  1+\frac{u}{{\bar{u}}^{2}\,\bar{z}%
}-\frac{u(1+u-z)}{{\left(  1-u-z\right)  }^{2}}\right)  \,-\nonumber\\
&  \left(  \frac{1}{2}-\frac{u^{2}}{(1-u-z)^{2}}-\frac{u}{2(1-u-z)}\right)
\ln\bar{z}+\nonumber\\
&  \left(  \frac{1}{2\bar{u}\bar{z}}-\frac{u}{(1-u-z)^{2}}-\frac{u}%
{2\bar {u}(1-u-z)}\right)  \frac{(1-\bar{u}\,z)}{\bar{u}}\ln(1-\bar
{u}\,z)~,
\end{align}

\begin{align}
JC2  &  =\frac{\pi^{2}}{24}\left(  \frac{4}{u\bar{z}}-3\right)  -\left(
\frac{1}{\bar{u}}-\frac{2-u}{2\bar{u}^{2}\bar{z}}\right)  \ln u+\left(
\frac{3z-1}{2uz}\right)  \ln\bar {z}-\nonumber\\
&  \left(  \frac{3}{2u}+\frac{2-u}{2\bar{u}^{2}\bar {z}}+\frac
{2u-1}{2uz\bar{u}^{2}}\right)  \ln(1-\bar {u}z)+\nonumber\\
&  \left(  1-\frac{1}{u\bar{u}\bar{z}}\right)  \text{Li}_{2}(\bar{u})+\left(
1-\frac{1}{u\bar{z}}\right)  \text{Li}_{2}(z)-\left(  1-\frac{1}{u\bar{u}%
\bar{z}}\right)  \text{Li}_{2}(z\bar{u})+\nonumber\\
&  \text{S}\left(  \frac{\bar{u}}{u}\right)  +\left(  \frac{1-z\bar {u}%
}{u~z}\right)  \left(  \text{S}\left(  \frac{z}{\bar{z}}\right)
-\text{S}\left(  \frac{z\bar{u}}{1-z\bar{u}}\right)  \right)
\end{align}

\begin{align}
JC3  &  =\mathrm{Re}JC3+i\pi\mathrm{Im}JC3,\\
\mathrm{Im}JC3  &  =\frac{3}{2}-\frac{u^{2}}{\,{\left(  1-u-z\right)  }^{2}%
}+\frac{u}{\,2{\left(  1-u-z\right)  }}-\frac{{u}^{3}\,}{{\left(
1-u-z\right)  }^{3}}\ln\left(  \frac{u}{\bar{z}}\right)  -\ln(u\,\bar{z})~,
\end{align}%
\begin{align}
\mathrm{Re}JC3  &  =1-\frac{7{\pi}^{2}}{12}-\frac{u^{3}\,\text{{I}}(u,\bar
{z})}{{\left(  1-u-z\right)  }^{2}}-\left(  \frac{3}{2}-\frac{u^{2}%
}{\,{\left(  1-u-z\right)  }^{2}}+\frac{u}{\,2{\left(  1-u-z\right)  }%
}\right)  \ln u+\nonumber\\
&  +\left(  3-4u-\frac{3}{z}+\frac{u^{2}\,\left(  z-1+3\,u\right)  }{{\left(
1-u-z\right)  }^{2}}\right)  \frac{\,\ln\bar{z}}{2u}+\frac{1}{2}\ln^{2}\left[
u\bar{z}\right]  +\nonumber\\
&  +\left(  -\frac{3\bar{u}}{2u}+\frac{3-4u}{2uz\bar{u}^2}-\frac{(2-u)u^{2}%
}{\bar{u}{\left(  1-u-z\right)  }^{2}}-\frac{2u(2u-1)-u^{3}}{2\bar{u}%
^{2}(1-u-z)}\right)  \ln\left[  1-\bar{u}z\right] \nonumber\\
&  \frac{\left(  1-\bar {u\,}z\right)  }{u\,z}\left(  \,S\left(  \frac
{z}{\bar{z}}\right)  -S\left(  \frac{\bar {u}\,z}{1-\bar {u}%
\,z}\right)  \right)  ~,
\end{align}

\begin{align*}
JC4  &  =\frac{1}{2}+\left(  \frac{2-u}{uz}-\frac{1}{2uz^{2}}-\frac{3-2u}{2u}\right)
\ln\bar{z}-\left(  \frac{3}{2uz}-\frac{1}{2uz^{2}\bar{u}}\right)  \left(
1-z\bar{u}\right)  \ln\left[  1-z\bar{u}\right]  -\\
&  \frac{\left(  1-z\bar{u}\right)  }{uz}\left(  \,\text{S}\left(  \frac
{z}{\bar{z}}\right)  -\text{S}\left(  \frac{\bar {u}\,z}{1-\bar
{u}\,z}\right)  \right)  ,
\end{align*}

\begin{equation}
JC6=\frac{\pi^{2}}{24\bar{u}}+\frac{3}{2\bar{u}}-\left(  \frac{1}{\bar{u}%
}+\frac{u}{2\bar{u}^{2}}\right)  \,\ln u+\frac{1}{\bar{u}}\,\text{S}\left(
\frac{\bar {u}}{u}\right)  ,
\end{equation}

\[
JC7=\frac{1}{2\bar{u}}+\frac{u}{2\bar{u}^{2}}\ \ln u,
\]

\begin{align}
JD1  &  =\mathrm{Re}JD1+i\pi\mathrm{Im}JD1,\\
\mathrm{Im}JD1  &  =-\frac{1}{u-z}+\frac{\bar{z}\,}{{\left(  u-z\right)  }%
^{2}}\ln\left(  \frac{\bar{z}}{\bar{u}}\right)  ~,\\
\mathrm{Re}JD1  &  =-\frac{\pi^{2}}{24\bar{u}}-\frac{\bar{z}\,\text{{I}}%
(\bar{u},\bar{z})}{u-z}+\left(  \frac{1}{u-z}-\frac{1}{u\,\bar{z}}\right)
\,\ln\bar{u}+
\frac{\ln\bar {z}}{u-z}-\frac{\left(  1-u\,z\right)  \,\ln
(1-u\,z)}{u\,\bar{z}\,\left(  u-z\right)  }-\frac{1}{\bar{u}}\text{S}\left(
\frac{u}{\bar{u}}\right)  ~,
\end{align}

\begin{align*}
JD2  &  =\frac{\pi^{2}}{24\bar{u}}\left(  \frac{z-5-3u\bar{z}}{\bar{z}%
}\right)  -\frac{\ln\bar{u}}{u\bar{z}}-\frac{\ln\bar{z}}{z\bar{u}}%
+\frac{(1-uz)}{uz\bar{u}\bar{z}}\ln\left[  1-uz\right]  +\\
&  \frac{1+u\bar{z}}{z\bar{u}}\left(  \text{Li}_{2}\left(  u\right)
+\text{Li}_{2}\left(  z\right)  -\text{Li}_{2}\left(  uz\right)  \right)
-\text{S}\left(  \frac{u}{\bar{u}}\right)  -\frac{1-uz}{z\bar{u}}\left(
\text{S}\left(  \frac{z}{\bar{z}}\right)  -\text{S}\left(  \frac{uz}%
{1-uz}\right)  \right)  ,
\end{align*}

\begin{align}
JD3  &  =\mathrm{Re}JD3+i\pi\mathrm{Im}JD3,\\
\mathrm{Im}JD3  &  =\frac{1}{2\,\bar{u}}+\frac{1}{u-z}+\left(  \frac{1}%
{\bar{u}}+\frac{\bar{u}}{(u-z)^{2}}+\frac{1}{u-z}\right)  \ln\bar
{u}+
\left(  2-\frac{1}{\bar{u}}-\frac{\bar{u}}{(u-z)^{2}}-\frac{1}{u-z}\right)
\ln\bar{z}
\\
\mathrm{Re}JD3  &  =\pi^{2}\left(  \frac{7}{12}+\frac{1}{24\bar{u}}\right)
+\left(  \frac{1-u\bar{u}}{\bar{u}}-\frac{uz}{\bar{u}}+\frac{\bar{u}}%
{u-z}\right)  \text{{I}}(\bar u,\bar z)-
\left(  \frac{1}{2\,\bar{u}}+\frac{1}{u-z}\right)  \ln\bar{u}+
\nonumber\\ &
\frac
{1-z {u}}{\bar{u}z(u-z)}\ln\left[  1-z {u}\right]  -
\left(  \frac{u}{2\bar{u}}+\frac{1}{\,\left(  u-z\right)  }+\frac{1}%
{\bar{u}\,z}\right)  \,\ln\bar{z}-\frac{1}{2}\ln^{2}\left[  \bar{z}\bar
{u}\right]  -
\nonumber\\  &
\frac{\left(  \bar{z}-uz\right)  \,}{\bar{u}\,z}\text{S}\left(  \frac
{z}{\bar{z}}\right)  +\frac{\left(  1-u\,z\right)  \,}{\bar{u}\,z}%
\text{S}\left(  \frac{u\,z}{1-u\,z}\right)  ,
\end{align}

\[
JD4=\frac{5}{2}+\frac{\bar{z}}{z}\ln\bar{z}+\frac{1-uz}{z\bar{u}}\left(
\text{S}\left(  \frac{z}{\bar{z}}\right)  -\text{S}\left(  \frac{u\,z}%
{1-u\,z}\right)  \right)  ,
\]

\begin{align}
JD6  &  =\mathrm{Re}JD6+i\pi\mathrm{Im}JD6~,\\
\mathrm{Im}JD6  &  =\frac{\bar{z}}{\,\bar{u}\,\left(  u-z\right)  }+\frac
{\bar{z}\,\left(  1-u\,\left(  \bar{u}+z\right)  \right)  }{\bar{u}\,{\left(
u-z\right)  }^{2}}\,\ln\left(  \frac{\bar{u}}{\bar{z}}\right) \\
\mathrm{Re}JD6  &  =\frac{9}{2\,\bar{u}}+\left(  \frac{1}{\bar{u}}-\frac
{uz}{\bar{u}}+\frac{\bar{u}}{u-z}\right)  \text{{I}}(\bar {u}%
,\bar z)-
\left(  \frac{1}{\bar{u}}+\frac{1}{u-z}\right)  \ln(\bar {u}%
\,\bar {z})+\frac{\,\left(  1-u\,z\right)  \,}{\,\bar{u}\,z(u-z)}%
\ln(1-u\,z),
\end{align}
where for brevity we used new notation $S(x)$ $:$%

\begin{equation}
S(x)=\frac{1}{2}\ln^{2}\left(  1+x\right)  +\int_{0}^{1}\frac{d\alpha}{\alpha
}\ln(1+x\alpha)=\text{Li}_{2}\left(  \frac{x}{1+x}\right)  +\ln^{2}\left(
1+x\right)  .
\end{equation}
All other functions have been defined in the text.

\end{document}